\pdfoutput=1
\documentclass[prd,aps,twocolumn,a4paper,showkeys,nofootinbib]{revtex4-1}

\usepackage{graphicx,psfrag}
\usepackage{mathrsfs}
\usepackage{amsmath,amsfonts,amssymb}
\usepackage{multirow}
\usepackage{comment}
\usepackage[normalem]{ulem}
\usepackage{xurl}
\usepackage[colorlinks=true, linkcolor=black]{hyperref}
\usepackage{enumitem}
\usepackage{tcolorbox}
\usepackage{subfiles}

\newcommand{\be}{\begin{equation}}
\newcommand{\ee}{\end{equation}}
\newcommand{\bea}{\begin{eqnarray}}
\newcommand{\eea}{\end{eqnarray}}
\newcommand{\bel}{\begin{align}}
\newcommand{\eel}{\end{align}}

\DeclareMathOperator{\const}{const}

\def\l{\ell}

\def\M{{\cal M}}
\def\B{{\cal B}}

\def\Msun{{\rm M_{\odot}}}

\def\GMc2{{\rm G M_{\odot} c^{-2}}}

\def\tLam{{\tilde\Lambda}}

\def\spin{{\boldsymbol \chi}}

\def\bajes{{\tt bajes}}

\def\teob{{\tt TEOBResumS}}
\def\teobspa{{\tt TEOBResumSPA}}
\def\mlgwbns{{\tt mlgw\_bns}}
\def\model{\mlgwbns}

\def\eg{{\it e.g.}~}
\def\ie{{\it i.e.}~}

\usepackage{pifont} 

\usepackage{color}
\definecolor{cyan}{rgb}{0,0.9,0.9}
\definecolor{orange}{rgb}{0.9,0.5,0}
\definecolor{magenta}{rgb}{1,0,1}
\definecolor{purple}{rgb}{0.8,0.4,0.8}
\definecolor{gray}{rgb}{0.8242,0.8242,0.8242}
\definecolor{light-gray}{gray}{0.95}

\newcommand{\mbversion}{\texttt{0.12.0}}

\usepackage[]{acro}

\DeclareAcronym{EOB}{
  short = EOB,
  long = Effective One Body
}
\DeclareAcronym{GW}{
  short = GW,
  long = gravitational wave
}
\DeclareAcronym{PN}{
  short = PN,
  long = Post-Newtonian
}
\DeclareAcronym{ROQ}{
  short = ROQ,
  long = Reduced Order Quadrature
}
\DeclareAcronym{BNS}{
  short = BNS,
  long = binary neutron star
}
\DeclareAcronym{BBH}{
  short = BBH,
  long = binary black hole
}
\DeclareAcronym{GR}{
  short = GR,
  long = General Relativity
}
\DeclareAcronym{PE}{
  short = PE,
  long = parameter estimation
}
\DeclareAcronym{SPA}{
  short = SPA,
  long = stationary phase approximation
}
\DeclareAcronym{ML}{
  short = ML,
  long = machine learning
}
\DeclareAcronym{PCA}{
  short = PCA,
  long = principal component analysis
}
\DeclareAcronym{PSD}{
  short = PSD,
  long = power spectral density
}
\DeclareAcronym{CBC}{
  short = CBC,
  long = compact binary coalescence
}
\DeclareAcronym{FFT}{
  short = FFT,
  long = fast Fourier transform
}
\DeclareAcronym{NR}{
  short = NR,
  long = Numerical Relativity
}

\usepackage{tikz}
\usetikzlibrary{shapes,arrows,shadows}

\begin{document}

\title{Combining effective-one-body accuracy and reduced-order-quadrature speed for
binary neutron star merger parameter estimation with machine learning
}

\author{Jacopo \surname{Tissino}$^{1,2}$}
\author{Gregorio \surname{Carullo}$^{3,4}$}
\author{Matteo \surname{Breschi}$^{3}$}
\author{Rossella \surname{Gamba}$^{3}$}
\author{Stefano \surname{Schmidt}$^{5,6}$}
\author{Sebastiano \surname{Bernuzzi}$^{3}$}

\affiliation{$^{1}$ Gran Sasso Science Institute (GSSI), I-67100 L'Aquila, Italy}
\affiliation{$^{2}$ INFN, Laboratori  Nazionali  del  Gran  Sasso,  I-67100  Assergi,  Italy}
\affiliation{$^{3}$ Theoretisch-Physikalisches Institut, Friedrich-Schiller-Universit{\"a}t Jena, 07743, Jena, Germany}
\affiliation{$^{4}$ Niels Bohr International Academy, Niels Bohr Institute, Blegdamsvej 17, 2100 Copenhagen, Denmark}
\affiliation{$^{5}$ Nikhef, Science Park 105, 1098 XG, Amsterdam, The Netherlands}
\affiliation{$^{6}$ Institute for Gravitational and Subatomic Physics (GRASP), Utrecht University, Princetonplein 1, 3584 CC Utrecht, The Netherlands}

\date{\today}

\begin{abstract}
We present \mlgwbns, a gravitational waveform surrogate which allows for a significant improvement in the generation speed of frequency-domain waveforms for binary neutron star mergers, at a negligible cost in accuracy.
This improvement is achieved by training a machine-learning model on a dataset of waveforms generated with an accurate but comparatively costlier approximant: the state-of-the-art effective-one-body model \teobspa. 
When coupled to a reduced-order scheme, \mlgwbns {} can accelerate waveform generation up to a factor of ${\sim} 35$, outperforming all other approximants of similar accuracy.
By analysing GW170817 in realistic parameter estimation settings with our scheme, we showcase an overall speedup against \teobspa {} greater than an order of magnitude. 
Our methodology will bear a significant impact on the scientific program of next generation detectors by allowing routine usage of accurate effective-one-body models.
\end{abstract}


\maketitle

\section{Introduction}
\label{sec:intro}

Bayesian analyses of \ac{GW} data from compact binary mergers rely on extensive explorations of the posterior probability distribution of detected signals~\cite{Aasi:2013jjl, Veitch:2014wba} and chiefly hinge on accurate waveform models.
The latter represent the prediction of a \ac{GW} signal originated from a system described by a certain set of parameters \(\theta \).
As the sampling of the posterior distribution for a single \ac{GW} event typically requires the generation of\ \ \({\gtrsim}10^{7}\) waveforms, speed in their generation is essential.
This is especially compelling in view of next generation (XG) \ac{GW} detectors, for which the rate of events will be dramatically higher than the current one. 
The in-band duration of signals will also increase and, due to their low mass, long signals emitted by \ac{BNS} mergers will be most significantly affected.
Reducing the computational cost for this class of signals, maximising the scientific output of future large-scale experiments, is a goal of the utmost importance: this is the focus of the present work.

For general relativistic waveform models, speed and accuracy are often at odds.
For example, very fast waveform generation can be obtained with analytical \ac{PN} approximants~\cite{Buonanno:2009zt, Blanchet:2013haa}, but such templates lack in accuracy and tend to bias \ac{PE}.
Indeed, Bayesian analyses of \ac{BNS} signals in XG detectors~\cite{Smith:2021bqc,Pratten:2019sed,Williams:2022vct} have been demonstrated only using phenomenological~\cite{Dietrich:2019kaq} or \ac{PN} approximants~\cite{Schmidt:2019wrl}.
These waveform templates include only partial physical information, and are therefore expected to strongly bias \ac{PE} with XG detectors~\cite{Gamba:2020wgg, Williams:2022vct}.
For example, phenomenological approximants model the effect of spin precession, but do not contain unequal-mass tidal corrections, with the binary matter effects entirely determined by one single effective tidal parameter.
\ac{PN} models are unreliable close to merger and are only available (in the frequency domain) for binaries with spins aligned with the orbital angular momentum.
Incorporating the whole physics content of advanced waveform models (including higher harmonics, precession, eccentricity, self-spin interactions, beyond leading order adiabatic electric and magnetic-type tidal effects, dynamical tides) will be key to avoid biases, but currently can only be accomplished at a great increase in computational cost.

In particular, here we are interested in leveraging the \ac{EOB} approach~\cite{Buonanno:1998gg, Buonanno:2000ef,Damour:2000we,Damour:2001tu,Damour:2008qf, Damour:2009kr, Damour:2014sva, Bohe:2016gbl, Nagar:2018zoe, Lackey:2018zvw}, one of the most accurate state-of-the art frameworks for waveform generators. 
In this framework, the Hamiltonian description of the two-body problem in \ac{GR} is mapped to an effective problem of a single body orbiting in a Kerr-like deformed metric. 
The effective metric potentials are determined by suitably resummed \ac{PN} expressions that make the model predictive in the fast-motion and strong-field merger regime.
Gravitational waveforms are natively generated in the time-domain using the solution of the EOB equations of motion and a particular factorized and resummed analytical expression of the multipolar \ac{PN} waveform~\cite{Damour:2008gu}. 
The \ac{EOB} approach has the advantage of being both accurate to Einstein's equations, and flexible to the addition of analytical (e.g. \ac{PN}) information.
The faithfulness of inspiral-merger-ringdown models is increased by suitably informing them with \ac{NR} data, see e.g.~Refs.~\cite{Riemenschneider:2021ppj,Albertini:2021tbt} for recent work targeted at XG detectors. 
Analogously, \ac{BNS} inspiral-merger waveforms are obtained by augmenting the effective interbinary potential and waveform multipoles with tidal terms~\cite{Flanagan:2007ix,Damour:2009wj,Bini:2012gu,Bernuzzi:2012ci,Bernuzzi:2015rla,Hinderer:2016eia,Steinhoff:2016rfi,Akcay:2018yyh}.
Full inspiral-merger-postmerger \ac{BNS} waveforms can be constructed by hybridising the model with \ac{NR}-informed post-merger models~\cite{Breschi:2019srl, Breschi:2022xnc}.

The practical usage of the above \ac{EOB} model is hampered by the requirement of numerically solving of an ODE system, which brings a constant-time overhead and constrains the maximum rate of waveform generation. 
For current \ac{BNS} analyses, generating the \(\gtrsim 10^7\) templates required easily takes several weeks of CPU time.
A crucial element to improve the EOB model efficiency is the post-adiabatic method, introduced in Ref.~\cite{Nagar:2018gnk} for the \teob~model. The post-adiabatic iterative method yields an efficient yet accurate approximation of the \ac{EOB} Hamiltonian flow, removing the need to solve the related ODE for all but the very last stages of the inspiral.
This technique provides a significant speed up (a factor 10 or more for typical \ac{BNS} signals in the LIGO-Virgo~\cite{TheLIGOScientific:2014jea, TheVirgo:2014hva} band), but it is currently applicable only to quasi-circular mergers.
To further optimise the waveform generation, a desiderable feature for a fast approximant is to yield waveforms in the frequency domain, since the likelihood takes a simple form in the Fourier space, when assuming a Gaussian and stationary noise background.
A time-domain approximant, such as the one mentioned above, needs to be Fourier-transformed before use, which typically entails a slow-down up to an order of magnitude.
For this reason, a \ac{SPA} was introduced within the \ac{EOB} model \teob, yielding a fast and accurate frequency-domain approximant called \teobspa~\cite{Gamba:2020ljo}, which has been successfully applied to the analyses of GW170817 and GW190425 data, see e.g. Refs~\cite{Gamba:2020ljo, Breschi:2021wzr}.

The evaluation of a frequency-domain waveform approximant typically scales as \(t _{\text{waveform}} \approx t _{\text{overhead}} + N _{\text{points}} t _{\text{point}}\), where \(N _{\text{points}}\) is the number of grid points it is evaluated at. 
The per-frequency-point time \(t _{\text{point}}\) is typically on the order of few hundreds of nanoseconds, cannot be reduced below the CPU clock speed times the number of floating point operations required, and varies much less than \(t _{\text{overhead}}\) across models.
Thanks to the combination of \ac{SPA} and the post-adiabatic approach, the overhead time for \teobspa {} has been reduced to only tens of milliseconds. The fundamental limitation in reducing this number is that, even when evaluating the waveform at few frequency points, the full Hamiltonian flow must still be computed in a complete radial grid.
In current \ac{BNS} analyses the second term is typically dominant.
This leads to \ac{PE} times on the order of a few days on a modern computer cluster, which is acceptable for current event rates, but will not be for XG detectors.
A key observation is that \acp{ROQ}~\cite{Field:2013cfa} techniques can decrease the required value of \(N _{\text{points}}\) so much that the linear term becomes negligible comparable to the constant one.
The driving requirement behind this work is therefore to build a model with a much lower \(t _{\text{overhead}}\) --- which leads to a significantly faster \ac{PE} if combined with \ac{ROQ} --- while remaining faithful to the predictions of \ac{EOB}: this will enable accurate analyses of data from XG detectors.

One of the most promising approaches to achieve the necessary increase in efficiency is template acceleration through \ac{ML}.
The last few years saw a sharp rise in studies on this topic, a review of which can be found in Ref.~\cite{Tiglio:2021ysj}.
Most of these efforts, however, focused on \ac{BBH} signals.
A pioneering study on this was the one of Ref.~\cite{Chua:2018woh}, which developed a neural network to compute the liner combination coefficients of a generic BBH represented on a basis of waveforms. 
Along the lines of this work, Ref.~\cite{Khan:2020fso} also reached high performance, while retaining large faithfulness, compared the training waveforms.
Reference \cite{Schmidt:2020yuu} used instead a \ac{PCA} to drastically reduce the number of basis functions required for waveform reconstruction, while Ref.~\cite{Barsotti:2021wks} applied automated learning to select the best performing regression scheme (although varying only the BBH mass ratio $q$) and Ref.~\cite{Thomas:2022rmc} extended the latter effort to spin-precessing signals.
Finally, Ref.~\cite{Liao:2021vec} used a deep generative model for waveform generation.
It is worth noting that the models listed above work in the time-domain: while this ensures a smooth (hence more easily learnable) physical representation of the \ac{BBH} signal, a Fourier transform is still required to use the model in \ac{PE} applications.
The literature is less rich in the field of binary neutron star (\ac{BNS}) modelling. Reference \cite{Lackey:2016krb} developed a non-spinning surrogate model in the time domain, and extended it to aligned-spin \ac{BNS} using Gaussian Process regression \cite{Lackey:2018zvw}. 
Their parameter space is the same one we use, as discussed later.
Subsequently, Ref.~\cite{Lackey:2018zvw} built a fast frequency-domain surrogate of 
the spin-aligned model {\tt SEOBNRv4T} \cite{Hinderer:2016eia, Steinhoff:2016rfi}, again using Gaussian process regression.

In this work, we boost the efficiency of accurate EOB models generation by introducing {\mlgwbns}, the first {\it frequency domain} \ac{BNS} surrogate model relying on a neural network.
The key advancements introduced by our training algorithm rely on a combination of data compression techniques and analytical knowledge exploitation.
Its salient characteristics are: training on \teobspa {} waveform residuals relative to a \ac{PN} baseline, multi-step waveform downsampling and a final \ac{PCA} compression stage.
This way, the neural network must only learn a much simplified relation between the \ac{BNS} parameters $\theta$ and a low-dimensional representation of the waveform, which allows it to be shallow, in the end significantly decreasing the waveform computational overhead.
Synergic usage of this model with \ac{ROQ} compression techniques allows us to showcase more than an order of magnitude improvement in the analysis of current \ac{BNS} signals.
Most importantly, our technique immediately provides speedups up to $\sim 10^3$ for wider bandwidth analyses, enabling future studies to systematically exploit highly accurate \ac{EOB} models in full Bayesian \ac{PE} analyses involving XG detectors.

This paper is organized as follows. 
In Sec.~\ref{sec:model}, we describe the details of our method, while Sec.~\ref{sec:performance} is devoted to the performance analysis of our model in terms of timing and accuracy.
The improved capabilities of the resulting model are illustrated in Sec.~\ref{sec:PE}, where we show the results of a realistic \ac{PE} analysis on the \ac{BNS} transient GW170817~\cite{TheLIGOScientific:2017qsa,LIGOScientific:2018mvr}, additionally making use of a reduced order quadrature scheme to fully exploit the potential of our technique. 
Sec.~\ref{sec:con} presents final remarks and future research directions.

\paragraph*{Software availability. ---} Our model is released within the public \texttt{python} package {\mlgwbns}, available at \href{https://pypi.org/project/mlgw-bns/}{pypi.org/project/mlgw-bns/}.
The description in this paper refers to version \mbversion.
The package contains both the trained model described here, which can be used to generate waveforms out-of-the-box, and the full functionalities required to train new models at will (e.g.~by using a different approximant than \teobspa, or different parameter ranges).
The training time and memory requirements are both relatively small: a model can easily be trained on a laptop in a few hours.
The software we developed to achieve the frequency compression applied in the \ac{PE} stage is available at: \href{https://github.com/GCArullo/JenpyROQ}{github.com/GCArullo/JenpyROQ}.

\paragraph*{Conventions. ---}

We work in geometric units, setting \(G = c = 1\).
The total binary mass is denoted as $M= m_1 + m_2$, the mass ratio as $q = m_1 /m_2 \ge 1$, and the symmetric mass ratio as $\nu = m_1 m_2 / M^2$.
The dimensionless spin vectors are denoted as $\spin_i$ for $i=1,2$ and the spin components aligned with the orbital angular momentum $\textbf{L}$ are labeled as $\chi_i = \spin_{i}\cdot \textbf{L} / |\textbf{L}|$. 
The effective spin parameter is defined as \(\chi _{\text{eff}} = (\chi_1 m_1 + \chi_2 m_2 ) / M\).
The quadrupolar tidal polarizability parameters are defined as $\Lambda_{i}=({2}/{3})\,k_{2,i}\,C_i^{-5}$ for $i=1,2$, where $k_{2,i}$ and $C_i$ are the second Love number and the compactness of the $i$-th star, respectively. The reduced tidal parameter,
\begin{align}
  \label{eq:LambdaT}
  \tilde\Lambda &= \frac{16}{13}
  \frac{(M_1 + 12 M_2) M_1^4}{M^5}\Lambda_1 + (1\leftrightarrow 2)\,,
\end{align}
determines tidal interactions at leading post-Newtonian-order~\cite{Favata:2013rwa,Damour:2012yf}.
Masses, spins, and tidal parameters are collectively called the \emph{intrinsic parameters} of a \ac{BNS} system, \ie $\theta_{\rm int} =\{M,q,\chi_{1},\chi_{2},\Lambda_1,\Lambda_2\}$. 
The location and orientation of the source are identified by the \emph{extrisinc parameters} $\theta_{\rm ext}=\{D_L, \iota , \alpha, \delta, \psi, t_c,\phi_c\}$, \ie luminosity distance $D_L$, inclination angle $\iota$, right ascension angle $\alpha$, declination angle $\delta$, polarization angle $\psi$, time of coalescence $t_c$, and phase at the merger $\phi_c$.\footnote{This is the full extrinsic parameter set required to reconstruct the \(h_{\mu \nu }\) tensor, but within \mlgwbns {} concretely the sky position parameters \(\alpha\), \(\delta\) and \(\psi\) are not accepted: the polarizations are returned in a frame located at Earth and aligned with the source.}

The frequency-domain waveform from a compact binary coalescence can in general be written as 
\begin{align} \label{eq:emission-mode-decomposition}
h_+ (f) - i h_\times (f) = \frac{1}{D_L} \sum _{l=2}^{\infty } \sum _{m=-\ell}^{\ell} h_{\ell m} (f)\, {}_{(-2)}Y_{\ell m} (\iota, \varphi )\, ,
\end{align}
where the functions \(_{(-2)}Y_{\ell m}\) are the spin-weighted spherical harmonics, given \eg by equations II.7 and II.8 of Ref.~\cite{Ajith:2007a}, while the complex functions \(h_{\ell m} (f)\) are the frequency-domain modes of the GW strain.

The discussion in this paper is restricted to the $(\l,m)=(2,2)$ mode;
focusing on it, the two \ac{GW} polarizations can be simply written as 
\begin{subequations} \label{eq:22-mode-polarizations}
  \begin{align}
h_+ (f) &= \frac{1}{D_L} \sqrt{ \frac{5}{4 \pi }} h_{22}  (f) \frac{\cos^2 \iota + 1}{2} \\
h_ \times (f) &= \frac{1}{D_L} \sqrt{ \frac{5}{4 \pi }} h_{22}  (f) \cos \iota \,.
\end{align}
\end{subequations}

A relevant scalar product in waveform space is the Wiener product 
\begin{align}\label{eq:Wiener_ip}
(a | b) := 4 \Re \int_0^{ \infty } \frac{a^*(f) b(f)}{S_n(f)} \text{d}f\,,
\end{align}
where \(S_n\) is the \ac{PSD} of a given detector.
Results shown in Sec.~\ref{sec:performance} are computed considering \(S_n\) to be the expected Einstein Telescope \ac{PSD}, ET-D~\cite{EinsteinTelescope:2011fda,Hild:2010id}. 

In terms of this, the optimal match (or {\it faithfulness}) between waveforms $a$ and $b$ is given by 
\begin{align}
\mathcal{F}(a, b) 
:= \max_{t_0, \phi_0 } \frac{(a|b)}{\sqrt{(a|a)(b|b)}}\,,
\end{align}
where the maximum is taken over all possible time and phase shifts \(t_0 \) and \(\phi_0\) between the two waveforms. 
The mismatch is then defined as $\bar{\mathcal{F}}(a, b) := 1 - \mathcal{F}(a, b)$.

\section{Model construction}
\label{sec:model}

\subsection{Overview}

\mlgwbns {} is a surrogate waveform approximant based on a neural network that learns the relation between five intrinsic parameters of a binary system --- mass ratio, dimensionless spins, and quadrupolar tidal polarizabilities, collectively denoted as $\theta =\{q,\chi_{1},\chi_{2},\Lambda_1,\Lambda_2\}$ --- and the corresponding frequency-domain waveform mode $h_{22}(f;\theta)$.
The binary mass $M$ is not included in $\theta$ since the nontrivial mass scale in the \ac{BNS} problem is fully included in the tidal polarizability parameters.\footnote{As in the scale-invariant binary black hole case, the waveform's frequency dependence is really on the mass-rescaled parameter \(Mf = GMf / c^3\), not on \(f\) alone.} 
Concretely, within \mlgwbns{} a fixed reference mass of \(M _{\text{ref}} = 2.8 M_{\odot}\) is chosen and waveforms for generic masses are generated by the appropriate rescaling of both the waveform's amplitude and frequency; this is described in more detail in Sec.~\ref{sec:frequencies}.
Similarly, the other extrinsic parameters \(\theta _{\text{ext}}\) can be neglected when constructing an approximant: the dependence on them can be included analytically in the likelihood after a waveform has been generated.

The driving idea behind \mlgwbns {} is to have the neural network be as shallow and small as possible while retaining reconstruction accuracy;
this is accomplished by reducing the dimensionality of the waveform's description.
The first step to this end is to make a training dataset of \emph{residuals} from an analytical \ac{PN} baseline, which means the network only has to learn information in the high-frequency region, where the two models differ: this is described in Sec.~\ref{sec:residuals}. 

We then employ a reduced frequency grid and perform a \ac{PCA} in order to decrease the dimensionality of each waveform's representation to about 30 floating point numbers; this is described in Sec.~\ref{sec:dimensionality-reduction}.

Finally, a neural network is trained to reconstruct the relation between 
the parameters \(\theta\) and the \(\sim 30\) principal components, as described in Sec.~\ref{sec:neural-network}.

The training datasets for all the aforementioned stages are generated by drawing from the same uniform distribution on the parameters, in the intervals:
$$
q \in [1, 2]\,,\ \  \Lambda _i \in [5, 5000]\,,\ \ \chi _i \in [-0.5, 0.5] \,.
$$
These ranges correspond to a realistic prior choice in GW analyses of \ac{BNS}.
The random number generator used for the extraction is deterministically re-seeded for every new dataset, 
in order to ensure reproducibility as well as independence of the datasets.

The frequency-domain waveforms currently learned by \mlgwbns {} are those generated by the state-of-the-art \ac{EOB} model 
\teobspa{}; these will be denoted by a subscript \(\text{EOB}\) in the following discussion.
We train with \teobspa{} frequency domain waveforms as opposed to Fourier transforms of time-domain \texttt{TEOBResumS} waveforms for a few reasons: the two are closer than the intrinsic accuracy of \texttt{TEOBResumS} (\(\mathcal{\bar{F}} \lesssim 5 \times 10^{-4}\)~\cite{Gamba:2020ljo}); the \ac{SPA} waveforms are much smoother than the ones calculated with a \ac{FFT}, and therefore easier to represent with small amounts of frequency points (see Sec.~\ref{sec:downsampling}); the \ac{SPA} waveforms can be natively evaluated at arbitrary frequencies, allowing us to never employ a uniform frequency grid.

\begin{figure*}[ht]
\centering
\pgfdeclarelayer{background}
\pgfdeclarelayer{foreground}
\pgfsetlayers{background,main,foreground}

\definecolor{myred}{HTML}{c16c5a}
\definecolor{myyellow}{HTML}{faeba2}
\definecolor{myorange}{HTML}{de8261}
\definecolor{myviolet}{rgb}{0.258234, 0.038571, 0.406485,}
\definecolor{wrongultramarine}{rgb}{0.07, 0.04, 0.56}
\tikzstyle{generator}=[draw, fill=myviolet!20, text width=5em, 
    text centered, minimum height=2.5em]
\tikzstyle{parameters}=[generator, fill=myred!45, rounded corners, text width=15em]
\tikzstyle{ann} = [above, text width=5em, text centered]
\tikzstyle{algo} = [generator, text width=15em, fill=white!40, 
    minimum height=6em, rounded corners]
\tikzstyle{trained} = [parameters, fill=myorange!60, ellipse, text width=2em]
\tikzstyle{arrow} = [ultra thick,->]

\def\blockdist{2}
\def\edgedist{2.5}

\begin{tikzpicture}[node distance=2cm]
\node (eob) [generator] {Effective One Body};
\node (pn) [generator, right of=eob, xshift=2.5cm, yshift=-2.5cm] {Post-Newtonian};
\node (pred) [generator, right of=eob, xshift=7cm] {predicted waveform};

\node (manage) [algo, below of=eob, yshift=-.7cm] {\textbf{Management}
\begin{itemize}
    \addtolength{\itemindent}{-.5cm}
    \item greedy downsampling (\ref{sec:downsampling})
\end{itemize}
};

\node (dimred) [algo, below of=manage, yshift=-1cm] {\textbf{Dimensionality reduction} 
\\ 
\begin{itemize}
    \addtolength{\itemindent}{-.5cm}
    \item residual calculation (\ref{sec:residuals})
    \item PCA training (\ref{sec:pca})
\end{itemize}};

\node (nn) [algo, below of=dimred, yshift=-1cm] {\textbf{Neural Network} 
\\ 
\begin{itemize}
    \addtolength{\itemindent}{-.5cm}
    \item optimization (\ref{sec:hyperparameter-optimization})
    \item NN training (\ref{sec:neural-network})
\end{itemize}};

\node (trainedpca) [trained, right of=dimred, xshift=2.5cm] {PCA};

\node (trainednn) [trained, right of=nn, xshift=2.5cm] {NN};

\node (manage2) [algo, right of=manage, xshift=7cm] {\textbf{Management}
\begin{itemize}
    \addtolength{\itemindent}{-.5cm}
    \item extrinsic parameter inclusion
    \item resampling to user grid
\end{itemize}
};

\node (dimred2) [algo, right of=dimred, xshift=7cm] {\textbf{Reconstruction}
\begin{itemize}
    \addtolength{\itemindent}{-.5cm}
    \item PCA reconstruction
    \item residual recombination
\end{itemize}
};

\node (nn2) [algo, right of=nn, xshift=7cm] {\textbf{Prediction}
\begin{itemize}
    \addtolength{\itemindent}{-.5cm}
    \item NN evaluation
\end{itemize}
};

\node (params) [parameters, below of=nn2, yshift=-.7cm] {user parameters \((\theta _{\text{int}}, \theta _{\text{ext}})\)};

\path (nn.south)+(0, -1cm) node (training) {\Large{Training}};
\path (params.south)+(0, -1cm) node (prediction) {\Large{Prediction}};

\draw [arrow] (eob) -- (manage);
\draw [arrow] (pn) -- (dimred);
\draw [arrow] (pn) -- (dimred2);
\draw [arrow] (manage) -- (dimred);
\draw [arrow] (dimred) -- (nn);
\draw [arrow] (nn2) -- (dimred2);
\draw [arrow] (dimred2) -- (manage2);
\draw [arrow] (manage2) -- (pred);
\draw [arrow] (params) -- (nn2);
\draw [arrow] (dimred) -- (trainedpca);
\draw [arrow] (trainedpca) -- (dimred2);
\draw [arrow] (nn) -- (trainednn);
\draw [arrow] (trainednn) -- (nn2);

\begin{pgfonlayer}{background}
    \path (manage.west |- manage.north)+(-0.4,0.4) node (a) {};
    \path (training.south -| nn.east)+(+0.4,-0.4) node (b) {};
    \path[fill=myyellow!20,rounded corners, draw=black!50, dashed]
        (a) rectangle (b);

    \path (manage2.west |- manage2.north)+(-0.4,0.4) node (a) {};
    \path (prediction.south -| nn2.east)+(+0.4,-0.4) node (b) {};
    \path[fill=myyellow!20,rounded corners, draw=black!50, dashed]
        (a) rectangle (b);

\end{pgfonlayer}
\end{tikzpicture}
\caption{Flowchart for the operation of \mlgwbns.}
\label{fig:flowchart}
\end{figure*}
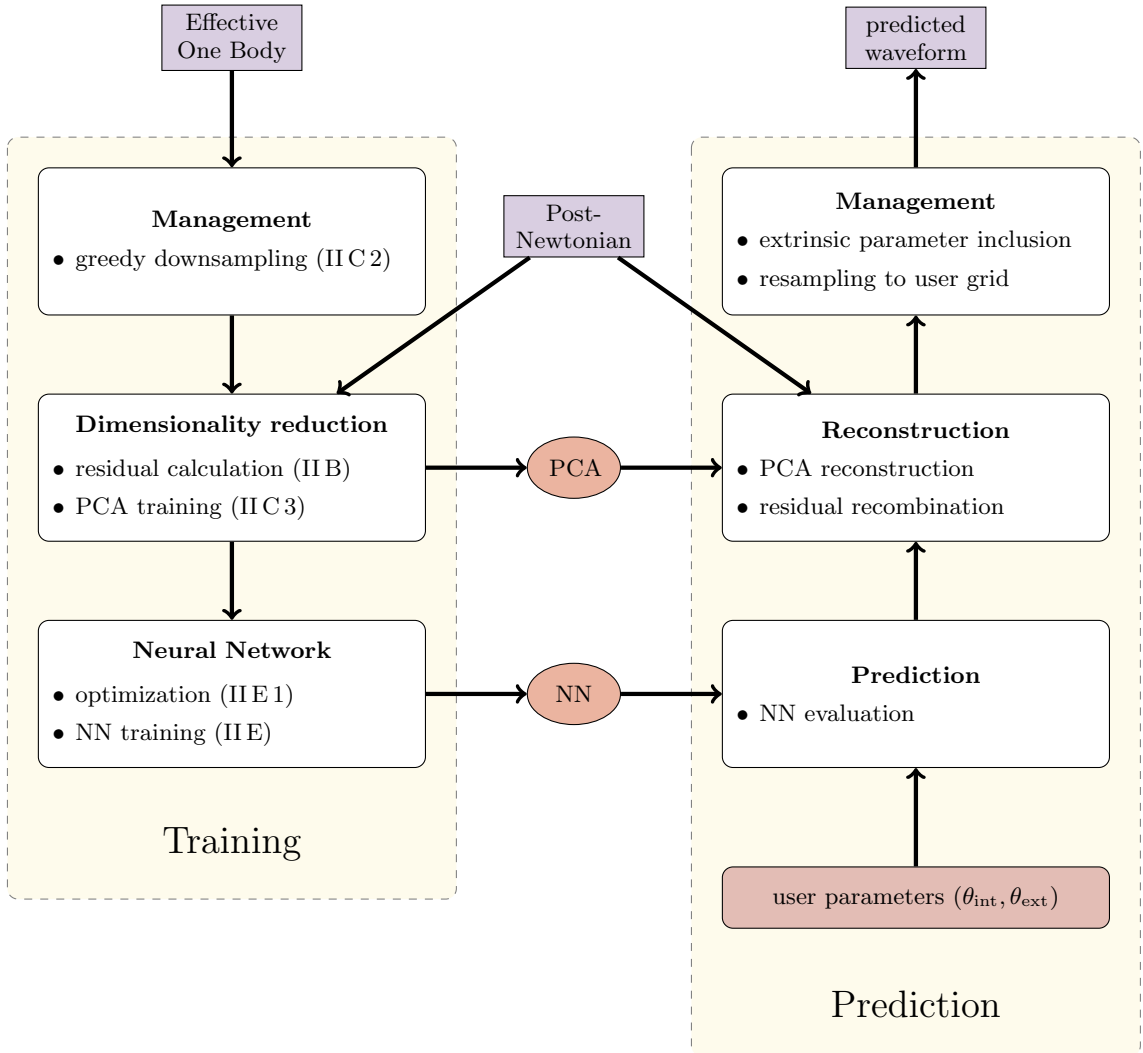

Figure~\ref*{fig:flowchart} shows a graphical outline of \mlgwbns's operation.

\subsection{Residuals from a Post-Newtonian baseline} \label{sec:residuals}

\begin{figure}[ht]
\centering
\label{fig:original-residuals}
\includegraphics[width=.48\textwidth]{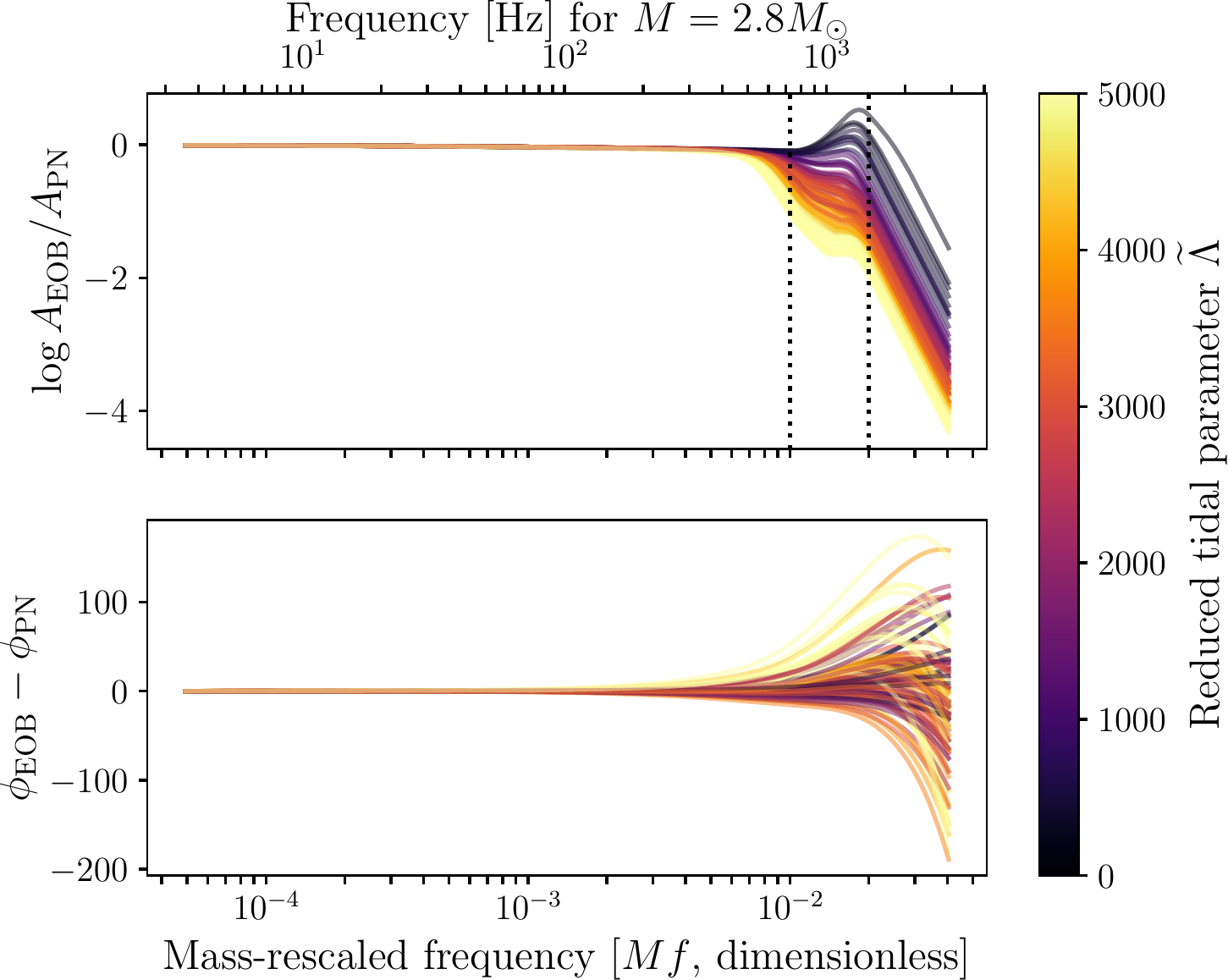}
\caption{Residuals of 100 \ac{EOB} waveforms to their \ac{PN} counterparts. The \ac{EOB} waveforms are chosen according to a uniform distribution in parameter space. 
}
\label{fig:original_residuals}
\end{figure}

We start with a polar representation of the waveform in amplitude and phase as 
\(h(f) = A _{\text{EOB}}(f) e^{- i \phi _{\text{EOB}} (f) }\).
Instead of reconstructing the waveform directly, \mlgwbns {} reconstructs its
residuals from a fiducial \ac{PN} model. 
The residuals are computed as 
\begin{subequations} \label{eq:amplitude-phase-residuals}
\begin{align}
\Delta A (f; \theta) &= \log \left(\frac{A _{\text{EOB}}(f; \theta)}{A _{\text{PN}}(f; \theta)}\right)
\\
\Delta \phi  (f; \theta) &= \phi _{\text{EOB}} (f; \theta) - \phi _{\text{PN}} (f; \theta)
\,.
\end{align}
\end{subequations}
and shown in Fig.~\ref{fig:original_residuals} for 100 sets of parameters.
The complete waveform is recovered from the predicted residuals 
\(\Delta A _{\text{pred}} (f; \theta ), \Delta \phi _{\text{pred}}(f; \theta )\)
as 
\begin{subequations} \label{eq:amplitude-phase-reconstruction}
\begin{align}
A _{\text{pred}}(f; \theta) &= A _{\text{PN}} (f; \theta) \exp( \Delta A _{\text{pred}} (f; \theta)) \\
\phi _{\text{pred}} (f; \theta) &= \phi _{\text{PN}}(f; \theta) + \Delta \phi _{\text{pred}}(f; \theta)
\,.
\end{align}
\end{subequations}
We use the \texttt{TaylorF2} approximant with 3.5PN-accurate amplitude, pseudo 5.5PN-accurate phase~\cite{Messina:2019uby} with 7.5PN-accurate tidal contributions~\cite{Damour:2012yf,Henry:2020ski} and the monopole-quadrupole 3PN contribution to the phase~\cite[Eqs.\ (50)--(52)]{Nagar:2018plt} (see also~\cite[Eq.\ (41)]{Nagar:2018zoe}).

The phase residuals computed as above typically exhibit large linear trends due to the different choices in the time-domain
alignment between the \ac{EOB} and \ac{PN} models (which corresponds to a linear phase term in the frequency domain).
These trends are not physically meaningful, but even small differences can result in a large effect: the variation over the whole frequency spectrum is of the order of \(2000 \text{Hz}\times 2 \pi \times \Delta t\) radians (for the reference mass), meaning that even single-millisecond shifts will yield tens of radians in difference. Typical shifts between the models used within \mlgwbns {} are of the order of tens of milliseconds, resulting in several hundreds of radians of meaningless phase difference. 
In order to remove this effect, the average slope \(\text{d} \Delta \phi / \text{d} f\) is first calculated between the first frequency sample and some higher frequency (typically chosen to be low enough to lie in the region of validity of the PN approximation) and then the corresponding linear term is subtracted from the residuals. Figure~\ref{eq:amplitude-phase-residuals} shows residuals with this procedure already applied.

This means that waveforms returned by \mlgwbns {} are aligned with the corresponding \ac{PN} ones, as opposed to the \ac{EOB} ones.
Since the prediction of the merger time within \mlgwbns {} is modelled on the \ac{EOB} one, this means that the predicted waveforms' mergers fluctuate by the same few tens of milliseconds.
This is inessential for the purposes of inspiral-only parameter estimation, but it can be problematic if we wish to extend the inspiral model with one for the post-merger~\cite{Breschi:2022xnc,Breschi:2022ens}.
A solution to this could be to reconstruct the time-shift dependence on the parameters \(\Delta t(\theta _{\text{int}})\), and de-shift the predicted waveforms after generating them with the \ac{PN} alignment; this is however not implemented in version \mbversion {} of \mlgwbns {} used in this work. 

\subsection{Dimensionality reduction} \label{sec:dimensionality-reduction}

Neural networks can be small and simple if the dimensionality of the data they must operate on is itself small.
Fortunately, the default representation of residuals (or waveforms) in frequency space contains a large amount of redundancy: this section discusses our approach to reducing the dimensionality of its representation.

The steps employed within \mlgwbns {} to this end are three: two of them are different techniques of decreasing the number of points in frequency space the residuals are sampled at, and the third is \ac{PCA}.

The orders of magnitude for how many floating point numbers are needed to represent waveforms or residuals starting at \(5 \text{Hz}\) after each of these steps are as follows (see also Figs.~\ref{fig:downsampling_comparison_amplitude} and~\ref{fig:downsampling_comparison_phase} for a breakdown of where these points are used in frequency space): 
\begin{enumerate}

  \item the default uniform frequency spacing requires \(\sim 2 \times 10^7\) points per waveform, scaling with \(f_0^{-8/3}\);
  
  \item the multibanding approach reduces this to \(\sim 2 \times 10^5\), scaling with \(f_0^{-5/3}\) \cite{Vinciguerra:2017ngf};
  
  \item the dataset-dependent greedy downsampling approach reduces this to \(\sim 3 \times 10^3\) for full waveforms or \(\sim 10^3\) for residuals;
  
  \item the \ac{PCA} representation, finally, only requires \(\sim 3 \times 10^1\) numbers per waveform.
  
\end{enumerate}

Uniform spacing is never used within \mlgwbns: a small number of waveforms is generated directly with the multibanded grid in order to train the greedy downsampling, and once this is done all further waveforms are generated on the smaller greedy downsampling grid.
This means that, even when starting from a very low initial frequency, we can easily work with a dataset of waveforms within the RAM of a laptop.

\subsubsection{Multibanding}
\label{sec:multibanding}

``Multibanding'' is the name we give to a technique of generating a frequency grid which is much smaller than the uniform one, not very dependent on the specifics of the dataset, and which may still be used to get a good representation of \ac{CBC} waveforms.

The starting point is the observation that a \ac{CBC} signal will always have a specific chirping profile, with high-frequency information only contained in a short (in time) section at the end.
The default frequency array used in signal processing, for a real-valued signal with duration \(T\) and time spacing \(\Delta t\), will be a uniform array from \(f = 0\) to \(f = 1 / 2 \Delta t\) (the Nyquist frequency), with spacing \(\Delta f = 1 / T\). 
As expected, this means there is no information loss: \(T / \Delta t\) real numbers are mapped to \((2 \Delta t )^{-1} / T^{-1}\) complex numbers.
This array describes high- and low-frequency information for all times: in the \ac{CBC} case this entails a lot of redundancy, since it is already known ahead of time that for the overwhelming majority of the signal there will be no high-frequency information.
We may construct a frequency array which is ``aware'' of this behavior~\cite{Vinciguerra:2017ngf,Smith:2016qas}. 
We start from the fact that the duration of a CBC signal starting from a frequency \(f_0\) is \(T \propto f_0^{-8/3}\), with a proportionality constant that can be analytically derived at Newtonian (0PN) order and which depends on the mass and the mass ratio~\cite{Maggiore:1900zz}: 
\begin{align}
T = \frac{5}{256} (\pi f_0)^{-8/3} M ^{-5/3} / \nu \,.
\end{align}
Then, we can make a frequency array for which the frequency spacing at each frequency is \(\Delta f (f) \approx 1 / T(f)\). 
This will mean we sample the low-frequency region much more finely than the high-frequency one, but locally each frequency band is described with the correct level of detail.

The approach used within \mlgwbns {} differs from the one used by Ref.~\cite{Vinciguerra:2017ngf} in two aspects.
First, whereas they approximate the uniformly-varying \(\Delta f\) by dividing the frequency domain into bands and using a different, uniform frequency for each of those, we construct a frequency array with continuously-varying spacing. Second, while they extend this sampling into the high-frequency regime, we use it only for frequencies lower than a certain pivot, typically \(f _{\text{pivot}} \approx 40 \text{Hz}\), while for higher frequencies we use uniform sampling. 
This is a conservative choice, motivated by the fact that at high frequency the 0PN expression for the time to merger cannot be expected to hold, combined with the fact that a uniform array with the spacing defined by \(\Delta f = 1/ T(40 \text{Hz}) \approx 0.02 \text{Hz}\) is not a large computational burden, resulting in only a few tens of thousands of points.
This approach, that we call \textit{multibanding}, needs to know something about the dataset: while the mass is kept fixed during the training, the mass ratio cannot be. 
The dependence is \(\Delta f \propto 1 / T \propto \nu \), and \(\nu \) scales inversely with the mass ratio \(q\) (which is \(>1\) here). 
Therefore, the \emph{smallest} \(\Delta f\) we should use as a lower bound corresponds to the largest \(q\) within the dataset; note, however, that this characteristic is shared by the uniform sampling, which is also defined by the three quantities \(f _{\text{min}}\), \(f _{\text{max}}\) and \(\Delta f\).

\begin{figure}[ht]
\centering
\includegraphics[width=.48\textwidth]{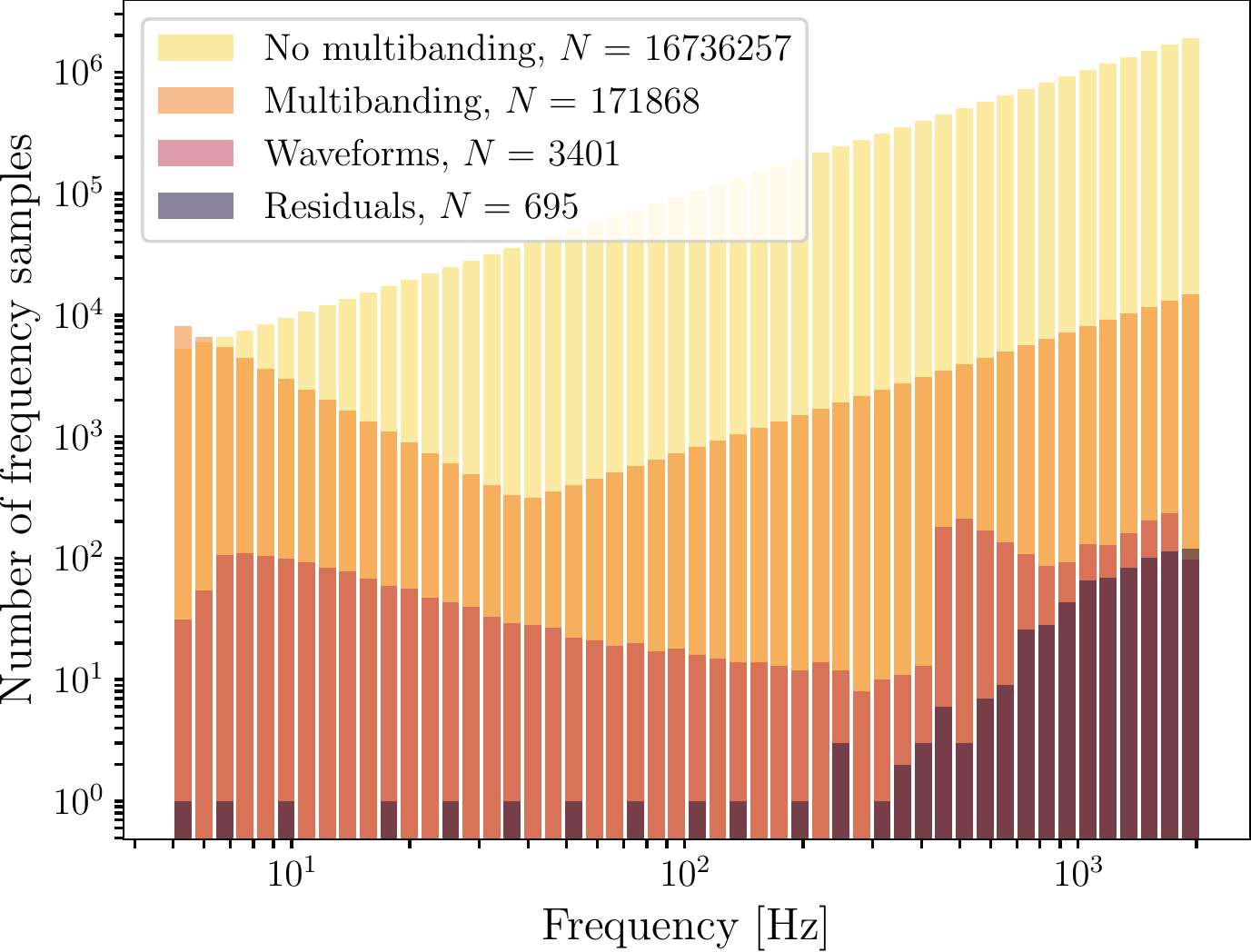}
\caption{Comparison of various ways to sample the amplitudes of a waveform. We show histograms of the arrays of frequencies used for the sampling, in the cases of no multibanding (uniform spacing \(\Delta f = \const\)), multibanding (discussed in Sec.~\ref{sec:multibanding}), and training the greedy algorithm discussed in Sec.~\ref{sec:downsampling} on 128 waveforms or 128 sets of residuals, computed as discussed in Sec.~\ref{sec:residuals}.}
\label{fig:downsampling_comparison_amplitude}
\end{figure}

\begin{figure}[ht]
  \centering
  \includegraphics[width=.48\textwidth]{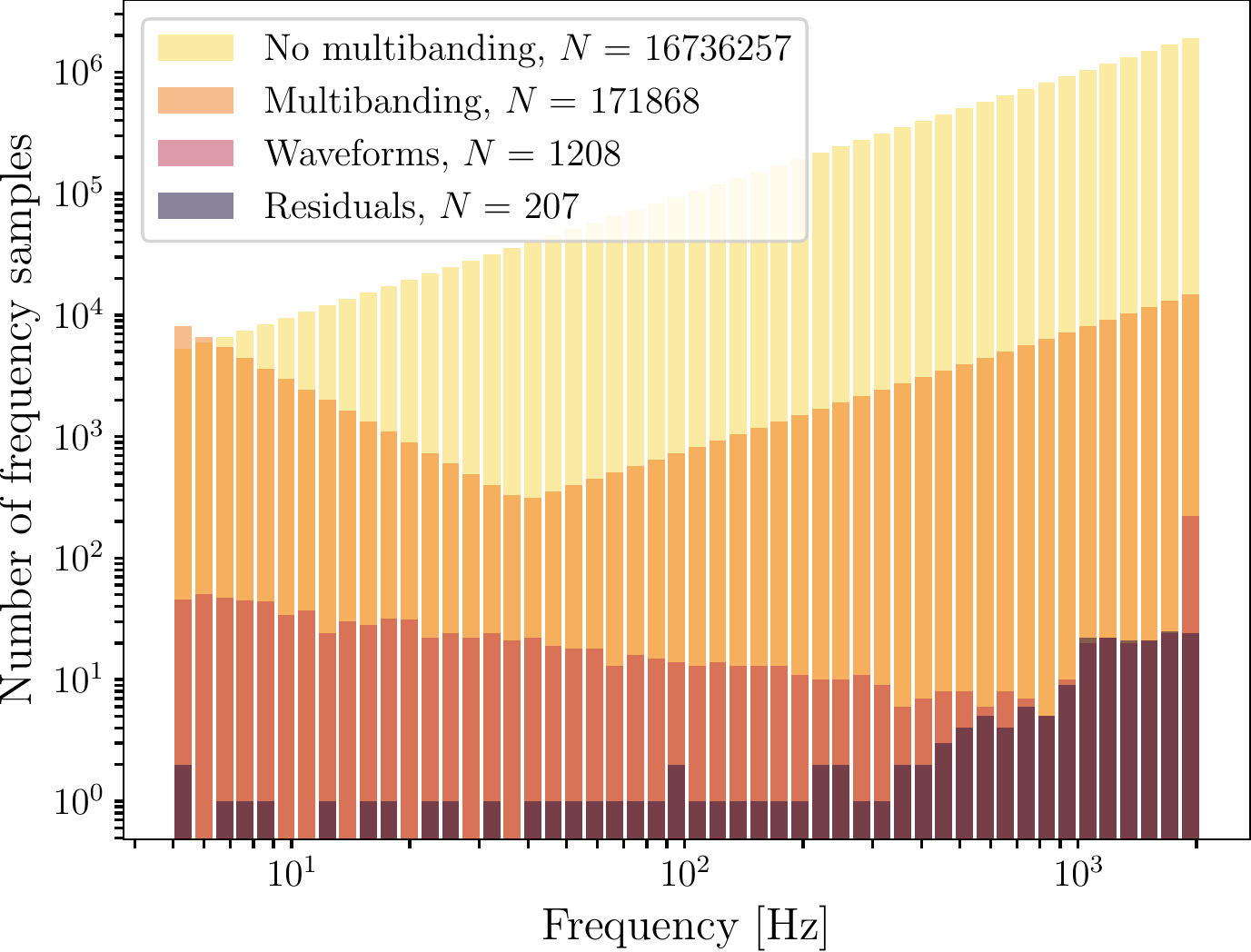}
  \caption{Same as Fig.~\ref{fig:downsampling_comparison_amplitude}, but training the greedy algorithm to reconstruct the phase of the same waveforms.}
  \label{fig:downsampling_comparison_phase}
\end{figure}

Figures~\ref{fig:downsampling_comparison_amplitude} and~\ref{fig:downsampling_comparison_phase} show histograms for the multibanding approach compared to the standard, uniform-in-frequency approach, for the case of waveforms starting at \(5 \text{Hz}\). 
The uniform-in-frequeny grid looks tilted in the histogram since the bins represent logarithmic frequency intervals, which increase in absolute width (\(\Delta f\)) as the frequency increases.
The general pattern to observe is that, as we make more and more assumptions about the waveforms we need to represent, the frequency array can shrink.
Multibanding is a rather safe choice, since it makes no more assumptions than uniform sampling, but it still provides at least an order-of-magnitude improvement in typical cases. 
The lower two histograms, labelled ``Waveforms'' and ``Residuals'', show the numbers of points that can be achieved when greedily selecting frequencies by requiring they allow us to reconstruct full EOB waveforms or their residuals (described in Sec.~\ref{sec:residuals}) respectively.

\subsubsection{Downsampling} \label{sec:downsampling}

While the multibanding reduces the size of the frequency arrays by orders of magnitude, especially for very low initial frequencies, we can do even better if we allow a heavier dependence on the specific dataset. 
Specifically, we can determine a set of points in frequency space such that any waveform in the dataset, if given at those points only, can be interpolated and retrieved at all frequencies within a certain accuracy. 

In order to achieve this goal, a greedy optimization technique is used. First, a set of waveforms is generated on the grid described in the previous section. These waveforms are then downsampled to a sparse grid, which can initially just consist of just the endpoints of the domain, and resampled with a cubic spline.\footnote{Cubic interpolation was found to be a good middle ground when accounting for computational complexity (which increases with interpolation order) and greedy grid size (which decreases with interpolation order).} 
The reconstruction error can then be measured for each of these waveforms: new points are added to the grid where it is worst. 
This procedure is iterated until all the given waveforms can be reconstructed within a certain tolerance, which we select to be \(10^{-5}\) for both amplitude and phase.
The downsampling is performed separately for amplitude and phase.

As the diagram in Fig.~\ref{fig:flowchart} shows, when reconstructing a waveform the ``residual recombination'' step happens before the ``resampling to user grid'' step.
This means that this downsampling procedure, which by itself is a generic algorithm, is applied to the \emph{full EOB waveforms} as opposed to the residuals described in Sec.~\ref{sec:residuals}.

While this requires us to use a slightly larger frequency grid (but still with \(< 10^4\) points), it was found to be generally faster than the alternative.

\subsubsection{Principal Component Analysis} \label{sec:pca}

Once the waveform has been downsampled, it is represented with \(n_A\) numbers for the amplitude and \(n_\phi \) for the phase.
Its dimensionality can be further reduced using \ac{PCA}. 
We collect all the residuals corresponding to each waveform in an array \(x = [\Delta A, \Delta \phi ] \in \mathbb{R}^{n_A + n_\phi }\) and construct a training dataset out of such arrays, \(\left\lbrace x_i \right\rbrace_i\), of which we may compute the mean \(\mu = \left\langle x \right\rangle\) and the covariance matrix 
\begin{align}
C = \left\langle (x - \mu ) (x- \mu )^{\top} \right\rangle \,.
\end{align}
This (symmetric, positive definite) matrix is diagonalized as \(C = V D V^{\top}\), where \(D = \operatorname{diag} (\lambda _i)\) is a diagonal matrix containing the eigenvalues of the covariance matrix, ordered so that \(\lambda _i \geq \lambda _{i+1}\).
The columns of \(V\) are the eigenvectors and, because of the ordering, the first \(k\) eigenvectors correspond to the \(k\) largest eigenvalues. 
Projecting a vector \(x\) onto the span of these \(k\) eigenvectors allows us to approximately represent it with only \(k\) numbers.
Specifically, if \(U\) is the \((n_A + n_\phi ) \times k\) submatrix  of \(V\) consisting of the \(k\) eigenvectors corresponding to 
the largest eigenvalues of the covariance matrix \(C\), we explicitly write the forwards and backwards transformations for \(x\) into its low-dimensional representation \(\widetilde{x}\): 
\begin{subequations} \label{eq:pca}
\begin{align}
x &\to \widetilde{x} =  U^{\top} (x - \mu )  \\
\widetilde{x} &\to x = U \widetilde{x} + \mu 
\,.
\end{align}
\end{subequations}

The number of principal components to keep can be tuned depending on the required final fidelity; including more of them increases the evaluation time of each waveform.

For simplicity, for the remainder of this work we always retain 30 principal components.
In principle this number could also be tuned, and its current value was mainly chosen to be ``safely large''.
This is confirmed by Fig.~\ref{fig:mismatches_by_n_train}: the reconstruction fidelity grows in a roughly linear fashion with the number of training points and its accuracy is never hampered by the number of \ac{PCA} up to fidelities \(\mathcal{\bar{F}} \lesssim 10^{-5}\).
As we will discuss in Sec.~\ref{subsec:roq-con}, even with this possibly suboptimal value of \(k\) our model is fast enough not to be the bottleneck in the evaluation of the likelihood. 

\subsection{Frequency band}\label{sec:frequencies}

As our detectors improve their sensitivity at low frequency, it is crucial to have a model which can be conveniently evaluated there.
In this section, we discuss the frequency band in which our model is trained, and how we may overcome the inherent limitation of only training down to a given frequency.

For the default model, which is provided with version \mbversion {} of \mlgwbns {} and whose performance is discussed in this work, 
the frequency range for which validity is guaranteed is \([5, 2048]\text{Hz}\), while the range of valid total masses is \([2, 4] M_{\odot}\): 
this means, as we shall discuss below, that the reference-mass model is trained in the range \(\approx [3.57, 2926] \text{Hz}\).

When the user requests frequencies within the training range, the model is able to directly yield a prediction; however this may be limiting, especially when considering multi-band observations.
The waveform at frequencies lower than the ones in the training range is well-described by the \ac{PN} approximation: therefore, waveforms predicted by \mlgwbns {} are natively hybridized with \ac{PN} ones at low frequency, as Sec.~\ref{sec:low-freq-bound} below describes.

\subsubsection{Mass rescaling} 

As mentioned in the introduction, we exclude the total mass \(M\) from the training parameters since the waveform only depends on the combination \(Mf\): this affects the frequency band in which we must train our model.

Suppose the user requires a waveform \(h(f; M, \theta )\) with total mass \(M\).
Then, the overall waveform is computed within \mlgwbns {} as
\begin{align}
h(f) = \frac{M}{ M _{\text{ref}}} h \left(\frac{fM}{M _{\text{ref}}}; M _{\text{ref}}, \theta\right)
\,,
\end{align}
which means that the user-given frequency grid will be shifted by a factor \(M / M _{\text{ref}}\).

In order for this to yield a valid waveform, however, the shifted frequencies must 
still lie within the model's training frequency range. 

Therefore, if we want our model to be applicable for all frequencies in a range \([f_1, f_2 ]\) and 
for all masses in a range \([M_1, M_2]\) we need to train the reference-mass model in a range 
\begin{align}
f \in \left[ f_1 \frac{M_1}{M _{\text{ref}}}, f_2 \frac{M_2 }{M _{\text{ref}}} \right]
\,.
\end{align}

\subsubsection{High frequency bound}

The model we are training on, \teobspa, describes the inspiral up to merger, which in the mass range of interest typically happens above \(2 \, \text{kHz}\).
After the merger, the remnant (a short- or long-lived neutron star, or a black hole) will emit a post-merger GW signal for which models exist~\cite{Clark:2015zxa,Breschi:2019srl,Easter:2020ifj,Soultanis:2021oia,Wijngaarden:2022sah,Breschi:2022xnc}, but which is considered separately from the \ac{EOB} waveform:
after the merger frequency, \teobspa {} waveforms are tapered with a powerlaw in the amplitude, \(A _{\text{EOB}} \propto f^{-10/3}\), and a linear relation in phase, \(\dot{\phi} _{\text{EOB}} = \dot{\phi} (f _{\text{max}})\), for \(f > f _{\text{max}}\) \cite[eqs.~S11-S12]{Gamba:2020ljo}.
This scaling is enforced as to ensure that the inverse Fourier transform of these waveforms is close to the time-domain waveform. 
Also, it means that the amplitude is guaranteed to remain positive (albeit quickly diminishing) at high frequency.

However, this implies an issue in the residuals computation of Eq.~\eqref{eq:amplitude-phase-residuals}: 
the baseline PN approximant is written as a power series in \(v = (\pi M f)^{1/3}\), 
which means that there is no guarantee that \( A _{\text{PN}}\) will remain positive in the high-frequency regime, and indeed
in practice, it often does become negative, which means that our residuals defined in Eq.~\ref{eq:amplitude-phase-residuals} diverge. 

We fix this by choosing a maximum frequency for the validity of the \ac{PN} model, and setting its amplitude to a constant value after that. 
This is not done ``sharply'', since that would propagate a discontinuity to the prediction: 
instead, we smoothly connect the expressions within an interval \([f_1, f_2] = [0.01/M, 0.02/M]\) as follows: 
for all \(f \in [f_1, f_2 ]\) we write 
\begin{align}
A _{\text{PN}}^{\text{new}} (f) =  \left(1 -  \zeta \left(x(f)\right)\right) A _{\text{PN}} (f) + \zeta \left(x(f)\right) C
\,,
\end{align}
where \(\zeta\colon [0, 1] \to [0, 1]\) is chosen so its derivative at the boundaries vanishes; specifically, we use 
\begin{align}
\zeta (x) = \frac{1}{2} \left(1 - \cos(\pi x)\right)
\,,
\end{align}
while 
\begin{align}
x(f) = \frac{f - f_1 }{f_2 - f_1}
\,.
\end{align}

The constant \(C\) is chosen to be equal to 20 in natural units; this is somewhat arbitrary, 
but it is roughly the value attained by \(A _{\text{EOB}}\) at \(f \sim 0.02/M\), 
as demonstrated by the first panel in Fig.~\ref{fig:original_residuals}: 
the value at \(Mf = 0.02\) is \(\log A _{\text{EOB}}(Mf=0.02) / C\), 
and one can see that it changes sign as we vary \(\widetilde{\Lambda}\). 

This shows that \(C=20\) is a reasonable middle ground for this parameter.
This choice will only have an impact on the network's ability to learn the residuals; 
if they are reconstructed correctly and the same modified \ac{PN} model is used both in training and reconstruction, 
the specifics of the modification do not matter, and the high-frequency continuation of our waveforms is equal 
to the \ac{EOB} one described at the beginning of this section.

For simplicity, for all frequencies higher than the maximum training one, we return a waveform which is identically equal to zero. 

\subsubsection{Low frequency bound} \label{sec:low-freq-bound}

For a typical \ac{BNS}, a frequency of \(5 \text{Hz}\) corresponds to about 2 hours before merger. 
This is close to the lower frequency limit for a ground-based detector, but 
for a multi-band observational campaign (including space- or Moon-based detectors) having a model able to 
be evaluated at arbitrarily low frequencies is very convenient.

The architecture in \mlgwbns {} makes this easily achievable: 
since we are reconstructing residuals from a \ac{PN} baseline, we may 
evaluate the waveform at arbitrarily low frequencies by setting the residuals to zero and just yielding the \ac{PN} waveform, which below \(5 \text{Hz}\) is a very good approximation of the true waveform:
as Fig.~\ref{fig:original_residuals} shows, the residuals approach 0 in the low-frequency regime.

For the phases, by subtracting an arbitrary linear term we can achieve \(\phi (f _{\text{min}}) = 0\) exactly, and \(\dot{\phi} (f _{\text{min}}) \approx 0\) to quite good accuracy, therefore we can simply yield \ac{PN} phases below \(f _{\text{min}}\) and our prediction above it.
For the amplitudes, this is not the case, and a discrepancy of the order of \(\Delta \log A \sim 5 \times 10^{-3}\) remains.

This discontinuity is fixed by a smoothing procedure: PN amplitudes corresponding to frequencies between \(f _{\text{min}} / 2\) and \(f _{\text{min}}\) are rescaled, so that the output of the model is 
\begin{equation}
  A(f) = 
\begin{cases}
  A _{\text{PN}} (f) & f < f _{\text{min}} / 2 \\
  A _{\text{PN}} (f) + \Delta A \zeta \left( \frac{2f}{f _{\text{min}}} - 1 \right) & f _{\text{min}} / 2 \leq f \leq f_{\text{min}} \\
  A _{\text{EOB}} (f)& f \geq f_{\text{min}}
\end{cases}
\end{equation}

\subsection{Neural Network} \label{sec:neural-network}

A feed-forward neural network is trained to reconstruct the map \(\theta \to \widetilde{x}\), where \(\theta\) is the vector of the 5 intrinsic parameters considered, while \(\widetilde{x}\) is a 30-dimensional \ac{PCA} representation of the residuals corresponding to the waveform generated by the \ac{EOB} model with the given parameters.

As our neural network we employ a \texttt{MLPRegressor} from the \texttt{scikit-learn} library~\cite{Pedregosa:2011bfd}, and the training is performed with the Adam algorithm for stochastic gradient descent~\cite{Kingma:2017rta}.

As it is common, the parameters \(\theta \) are rescaled to have mean \(0\) and standard deviation \(1\).
After the \ac{PCA} reduction, each component in the vector \(\widetilde{x}\) natively has comparable variance, but we may arbitrarily rescale them, which is equivalent to rescaling the eigenvectors in the matrix \(U\) defined in Sec.~\ref{sec:pca}.
Also, we know that the eigenvectors corresponding to the largest eigenvalues \(\lambda_i\) ``matter more'', in that they explain more variance. 
Therefore, as a preprocessing step we introduce a fixed rescaling of the vector \(\widetilde{x}\), as \(\widetilde{x}_i \to \widetilde{x}_i \lambda_i^\alpha\) for some tunable choice of \(\alpha \geq 0\).
The distance used during the training is then simply the Euclidean one between these rescaled \(\widetilde{x}\).

\subsubsection{Hyperparameter optimization} \label{sec:hyperparameter-optimization}

Several hyperparameters, which determine the network's properties and performance, must be chosen before training, such as
the number and size of hidden layers in the network, the activation function, the conditions for the termination of the training, the coefficient for the regularization term, and the coefficient \(\alpha\) defined above.
For a complete list, see App.~\ref{sec:appendix_B}, which details all the hyperparameters used within the default network discussed here.

The optimal set of hyperparameters may vary as the number of training waveforms used to train the network may change.
Heuristically, we might imagine that a complex network with many layers would be the best choice 
with many thousands of training waveforms, while it would overfit when using only a hundred waveforms for the training, 
for which the optimal configuration would be a smaller network.
The specific dependence of the reconstruction efficiency on these parameters is, however, high-dimensional and hard to explore since evaluating each point requires us to train the whole network.

We evaluate each possible set of hyperparameters by computing its average reconstruction error on a validation dataset, generated independently but from the same distribution as the training dataset; the reconstruction error is measured as the distance defined by
\begin{align} \label{eq:reconstruction-error}
\operatorname{dist}^{2}(\widetilde{x}_{\text{orig}}, \widetilde{x} _{\text{pred}}) = \frac{\lVert x _{\text{orig}}-x _{\text{pred}} \rVert^2}{n_\phi + n_A}
\,,
\end{align}
where \(n_\phi + n_A\) is the dimensionality of the vector \(x\), as defined in Sec.~\ref{sec:pca}: the distance is written in terms of the vectors \(x = [\Delta A, \Delta \phi ]\), 
reconstructed from the PCA-reduced \(\widetilde{x} \) predicted by the network.

The hyperparameters are optimised with the \texttt{optuna} package~\cite{Akiba:2019gsa} using a multi-objective 
tree-structured Parzen estimator~\cite{Ozaki:2020nor}, where the two cost functions being simultaneously optimised are 
\begin{enumerate}

    \item the average reconstruction accuracy on a validation dataset measured as in Eq.~\eqref{eq:reconstruction-error};
    
    \item the estimated time required for the generation of the training waveforms, 
    quantified by \(100 \text{ms}\) times the number of training waveforms, plus the 
    time needed to train the network.
    
\end{enumerate}

The training and validation datasets are randomized in each iteration.

The use of these two ``opposed'' cost functions allows for a Pareto front of optimal parameters to be computed.
This is a collection of parameter sets corresponding to different training dataset sizes; 
once this optimization has been run, for any given dataset size we have a set of good hyperparameters to train the network.

\begin{figure}[ht]
\centering
\includegraphics[width=.48\textwidth]{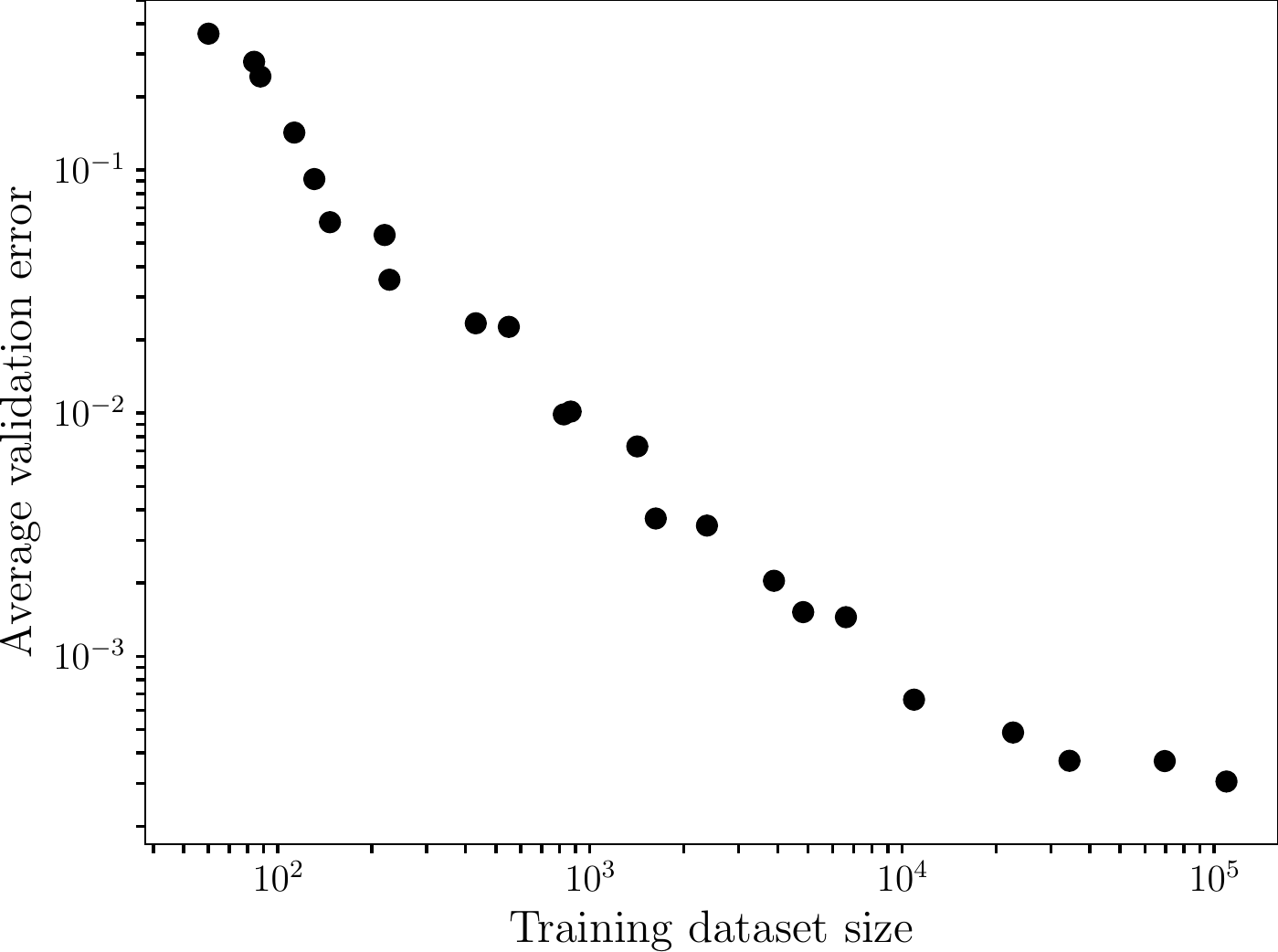}
\caption{Pareto front for the hyperparameter optimization. The vertical axis shows the average error, computed as in Eq.~\eqref{eq:reconstruction-error}. The flattening observed at large training dataset sizes is not necessarily real: computational constraints prevented a large amount of trials to be performed in that region.}
\label{fig:pareto_front}
\end{figure}

Such a collection --- with dataset sizes ranging from 50 to \(10^5\) training waveforms --- is provided with version \mbversion {} of \mlgwbns, and Fig.~\ref{fig:pareto_front} shows the validation errors as a function of training dataset size.
When creating a new model, a lookup may then be performed to recover the locally optimal hyperparameters for the amount of data available to the model.
This is efficient since it allows us to train new networks without re-running the optimization when the parameter 
space utilized remains relatively similar to the one used during the optimization procedure; 
we have however found that with significant changes to the parameter space (\eg including versus not including spin) 
the optimization had to be re-run since it was giving suboptimal results.

\section{Model performance}
\label{sec:performance}

\subsection{Accuracy}

\begin{figure}[ht]
\centering
\includegraphics[width=.48\textwidth]{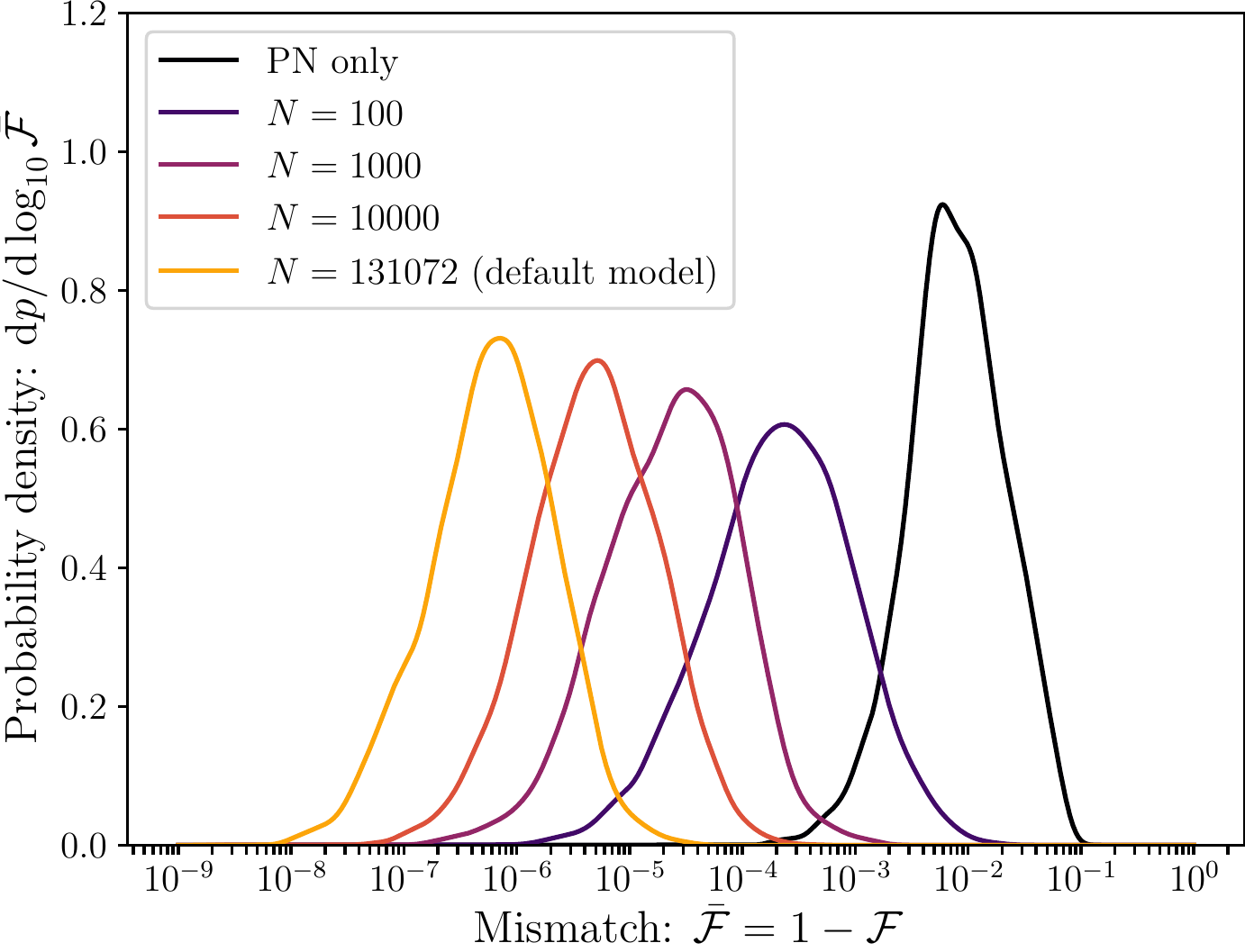}
\caption{Kernel Density Estimate representation of the mismatches 
between the waveforms reconstructed by \mlgwbns {} and the corresponding ones 
generated by the reference waveform generator, \teobspa, for uniformly-distributed sets of parameters \(\theta _{\text{int}}\) in the training ranges, and with constant total mass \(M = M _{\text{ref}} = 2.8M_{\odot}\). 
The curve labelled as ``PN only'' is obtained by comparing the baseline PN waveforms with the corresponding EOB ones, \ie setting the reconstructed residuals to zero; for the other curves we use the number indicated for both the training of the 
\ac{PCA} and for the training of the network, so the overall number of waveforms used is twice \(N\).
The same 4096 validation waveforms are used to generate each curve.
The mismatch is computed according to the predicted Einstein Telescope PSD, ET-D \cite{EinsteinTelescope:2011fda,Hild:2010id}, within the band \([3.57, 2926]\text{Hz}\) (see section \ref{sec:frequencies}).}
\label{fig:mismatches_by_n_train}
\end{figure}

Figure \ref{fig:mismatches_by_n_train} shows the mismatches between the reconstructed waveforms and the corresponding EOB ones. The mismatches are computed on validation datasets generated with the same distribution as the training ones, but with differently-seeded random number generators. The mismatches are computed according to the predicted Einstein Telescope PSD, ET-D \cite{EinsteinTelescope:2011fda,Hild:2010id}.
As shown by the figure, the accuracy measured through the mismatch \(\bar{\mathcal{F}}\) exhibits a roughly linear behavior \(\bar{\mathcal{F}} \sim 1 / N _{\text{train}} \). 

\begin{figure}[ht]
\centering
\includegraphics[width=.48\textwidth]{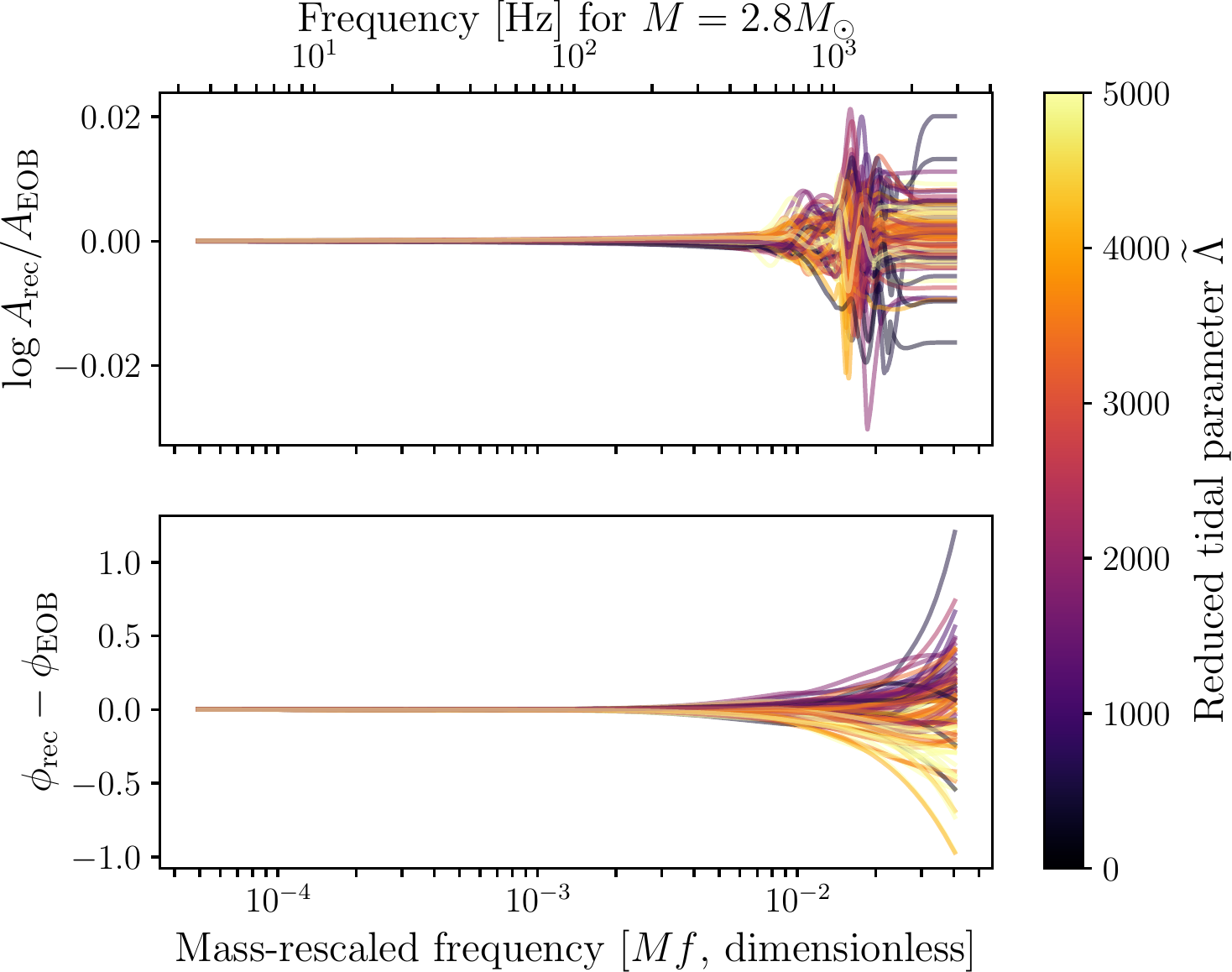}
\caption{Residuals of 100 reconstructed waveforms to the reference EOB ones. The parameters for them are uniformly distributed.
}
\label{fig:reconstruction-residuals}
\end{figure}

The reconstructed residuals corresponding to the best model of Fig.~\ref{fig:mismatches_by_n_train} (trained with \(2^{17} = 131072\) waveforms) are shown in Fig.~\ref{fig:reconstruction-residuals}. 
As one might expect, the residuals significantly differ from zero only in the high-frequency region, like the original residuals.
When considering the magnitude of the phase residuals, note that the logarithmic frequency axis distorts what may be linear trends: the temporal alignment chosen in the plot was not optimised to correspond to the best-match one, but instead to align the waveforms at low frequency.

\subsection{Speed}

\begin{figure}[ht]
\centering
\includegraphics[width=.48\textwidth]{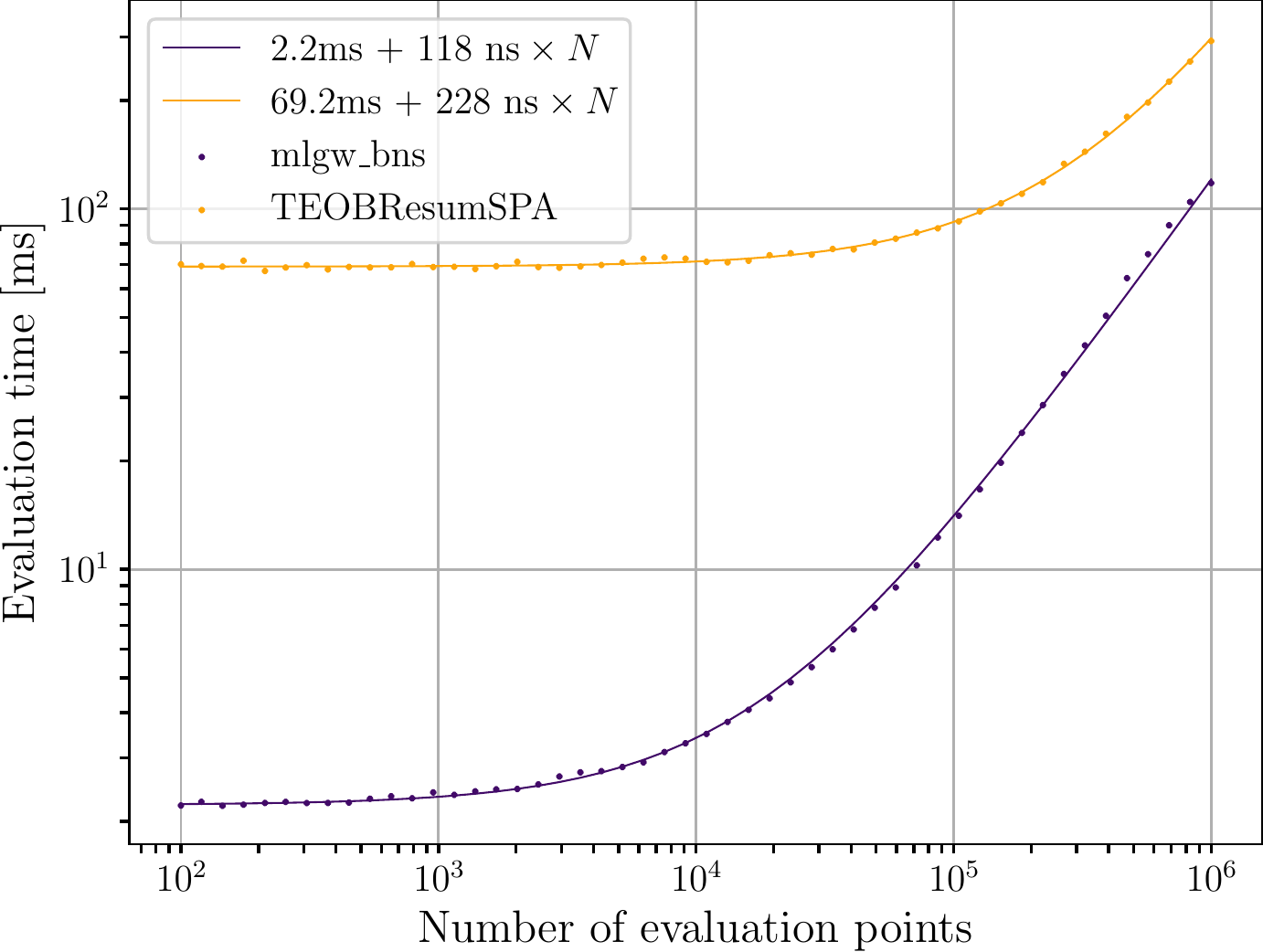}
\caption{Benchmarks of the evaluation time required for one waveform, 
with \teobspa {} and with \mlgwbns. Details on the benchmarking procedure can be found in section \ref{sec:appendix_C}.
For both approximants, we also show a fit with a model \(t = t_{\text{o}} + t_{\text{p}} N\).}
\label{fig:benchmarking-evaluation}
\end{figure}

The evaluation times for \mlgwbns {} are shown in Fig.~\ref{fig:benchmarking-evaluation} and compared to the evaluation times of \teobspa{}.
The significant acceleration provided by \mlgwbns {} is maximised when using compressed frequency grids with a small number of points, where it attains speedup factors of $\sim 35$.
The high template efficiency allows \mlgwbns {} to outperform not only \teobspa{} in waveform generation, but also all other state-of-the-art \ac{EOB} surrogate models available, as demonstrated in App.~\ref{sec:appendix_C}.

Both templates exhibit a similar behavior in the number of sampling points: \(t(N _{\text{sample}}) \sim t_{\text{o}} + t_{\text{p}} N _{\text{sample}}\). 
There is an approximately constant cost to evaluate the waveforms at small values of $N _{\text{sample}}$, while for large $N _{\text{sample}}$ the evaluation time scales linearly. 
This is due to the fact that, for both templates, there are operations that are approximately independent on the number of evaluation points. 
For \mlgwbns, these are running the parameters through the neural network and recomposing the result through \ac{PCA}. 
For \teobspa, the solution of the Hamiltonian flow using the post-adiabatic EOB iteration (at fixed number of points) and the subsequent ODE evolution for the last few orbits before merger \cite{Nagar:2018gnk}.
The linear regime is instead, for both templates, caused by the time to interpolate the waveform to each of the finely-spaced user-given frequency points, and performing other linear-time operations such as combining amplitude and phase into the Cartesian representation of the waveform. 

The linear-time operations taken by the two approximants are comparable; \teobspa {} is implemented in C and \mlgwbns {} in \texttt{python}, but several components in the latter are just-in-time compiled thanks to \texttt{numba} \cite{Lam:2015a}.
While the constant \(c_2\) might be whittled down by optimizing the implementation, the linear term can not be completely removed --- the program will have to do at least a few floating point operations for each point we are resampling at.
Therefore, if we want fast waveform evaluation it is important to use as small a number of points as we can, while retaining the desired accuracy.

Several approaches have been suggested towards this goal for \ac{PE} purposes: 
the simpler ones are similar in spirit to what has been discussed in Sec.~\ref{sec:multibanding}, 
using a smart coarser sampling than what the ``natural'' FFT grid would be.
More sophisticated approaches can be \ac{ROQ}s (discussed below in the context of \ac{PE}) or relative binning \cite{Zackay:2018qdy, Leslie:2021ssu}.

In Tab.~\ref{tab:wf_timings} we show a breakdown of the use of time within an evaluation of \mlgwbns, in the case of 1000 grid points.

\begin{table}[t]
\caption{Timing breakdown for the evaluation of a waveform on 1000 grid points with \mlgwbns. Values will fluctuate across evaluations, this table is only meant to be indicative of the ratios between them.}
\begin{tabular}{lclc}
\hline
\hline
Task & Time [\(\mu\)s] & Subtask & Time [\(\mu\)s]  \\
\hline
\multirow{2}{*}{Resampling} & \multirow{2}{*}{841} & Spline creation & 728 \\
&& Spline evaluation & 113 \\
\hline
\multirow{2}{*}{PN evaluation} & \multirow{2}{*}{653} & Amplitude & 434 \\
& & Phase & 219 \\
\hline
\multirow{3}{*}{PCA+NN} & \multirow{3}{*}{397} & NN & 326 \\
& & PCA & 41 \\ 
& & Misc. & 30 \\
\hline 
\multirow{3}{*}{Postprocessing} & \multirow{3}{*}{289} & Include extrinsic & 157 \\
&& Compute \(h = A e^{-i \phi }\) & 40 \\
&& Misc. & 90 \\
\hline\hline
Total & 2180 &&
\end{tabular}
\label{tab:wf_timings}
\end{table}

\section{Parameter estimation}
\label{sec:PE}

To showcase the benefits brought by our model in a realistic setting, we perform PE studies on the binary neutron star (\ac{BNS}) transient 
GW170817~\cite{TheLIGOScientific:2017qsa,LIGOScientific:2018mvr}.
In Sec.~\ref{subsec:teob-mlgw}, we first perform a full-scale validation, showing the compatible results of \ac{GW} inference using {\mlgwbns}, compared to the ones obtained with {\teobspa}.
Then, in Sec.~\ref{subsec:roq-con} we discuss and apply compression techniques capable of reducing the number of frequency nodes on which {\mlgwbns} needs to be evaluated for PE purposes. This step allows to fully exploit the benefits of our model, which displays the largest gain compared to {\teobspa} for a smaller number of frequency nodes (see Fig.~\ref{fig:benchmarking-evaluation}).
Sec.~\ref{subsec:roq-pe} finally repeats the \ac{PE} analysis combining {\mlgwbns} and such compression methods, showcasing more than order of magnitude speed gain obtainable with our \ac{ML} technique against {\teobspa} in a full-fledged \ac{PE} analysis.

In particular, we analyse the (deglitched) GWOSC data of LIGO and Virgo centered around GPS time 1187008857 
with a sampling rate of 4096~Hz and a duration of 128~s, considering the frequency range from [23, 2000]~Hz.
Our \ac{PE} relies on the MPI-parallelized {\bajes} pipeline~\cite{Breschi:2021wzr} and the 
${\tt dynesty}$ \cite{Speagle:2020} nested sampler.
The reported errors correspond to the 90\% confidence intervals and the $\log$ symbol refers to the natural logarithm.
The mass prior is chosen to be flat in the mass components $m_{1,2}$, although the 
sampling is then performed in $(\M,q)$, with ranges wide enough to capture the full posterior width.
We sample on aligned-spin components, with an isotropic prior bounded by $\chi_{1,2}\le 0.5$.
The prior on the tidal parameters is uniform in the ranges $\Lambda_{1,2}\in[5,5000]$ 
and the luminosity distance employs a volumetric prior in $D_L\in[1,75]~{\rm Mpc}$.
Other priors are set according to standard prescriptions in \ac{GW} astronomy~\cite{Breschi:2021wzr}.
We do not assume prior knowledge on electromagnetic counterparts.
We include spectral calibration envelopes with 10 logarithmic-spaced nodes for each detector.
For an overview of Bayesian inference of \ac{GW} signals see Refs.~\cite{Veitch:2009hd,Veitch:2014wba,thrane_talbot_2019,Breschi:2021wzr}.

\subsection{Full grid {\model} -- {\teobspa} comparison}\label{subsec:teob-mlgw}

\begin{figure*}[ht]
\centering
\includegraphics[width=0.96\textwidth]{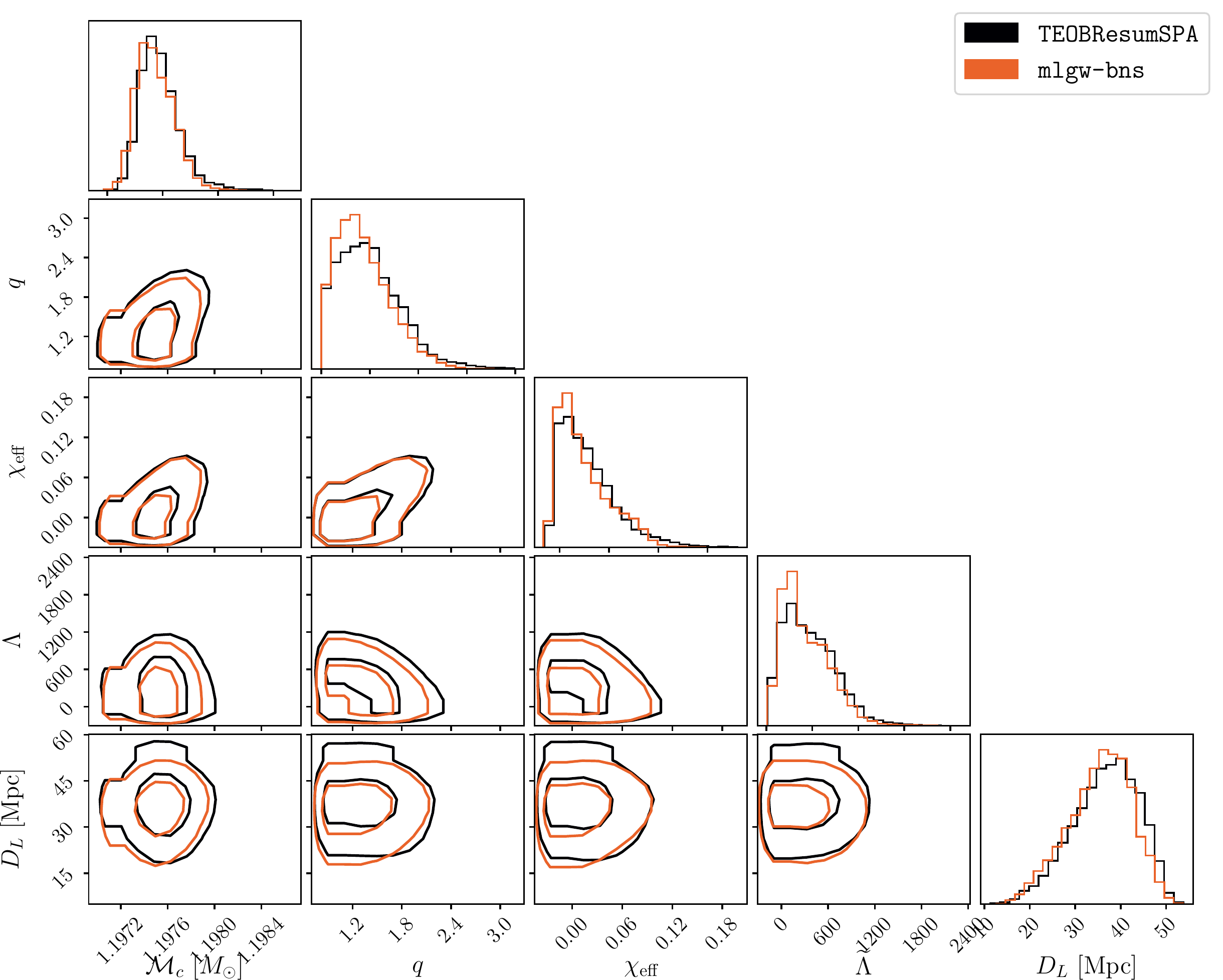}
\caption{Corner plot the posterior distribution for selected parameters reconstructed for GW170817, with \mlgwbns {} (orange) and \teobspa (black).
	The contours report the 50\% and the 90\% credibility regions. 
}
\label{fig:corner_posterior_mb_teob}
\end{figure*}

\begin{figure*}[ht]
\centering
\includegraphics[width=0.96\textwidth]{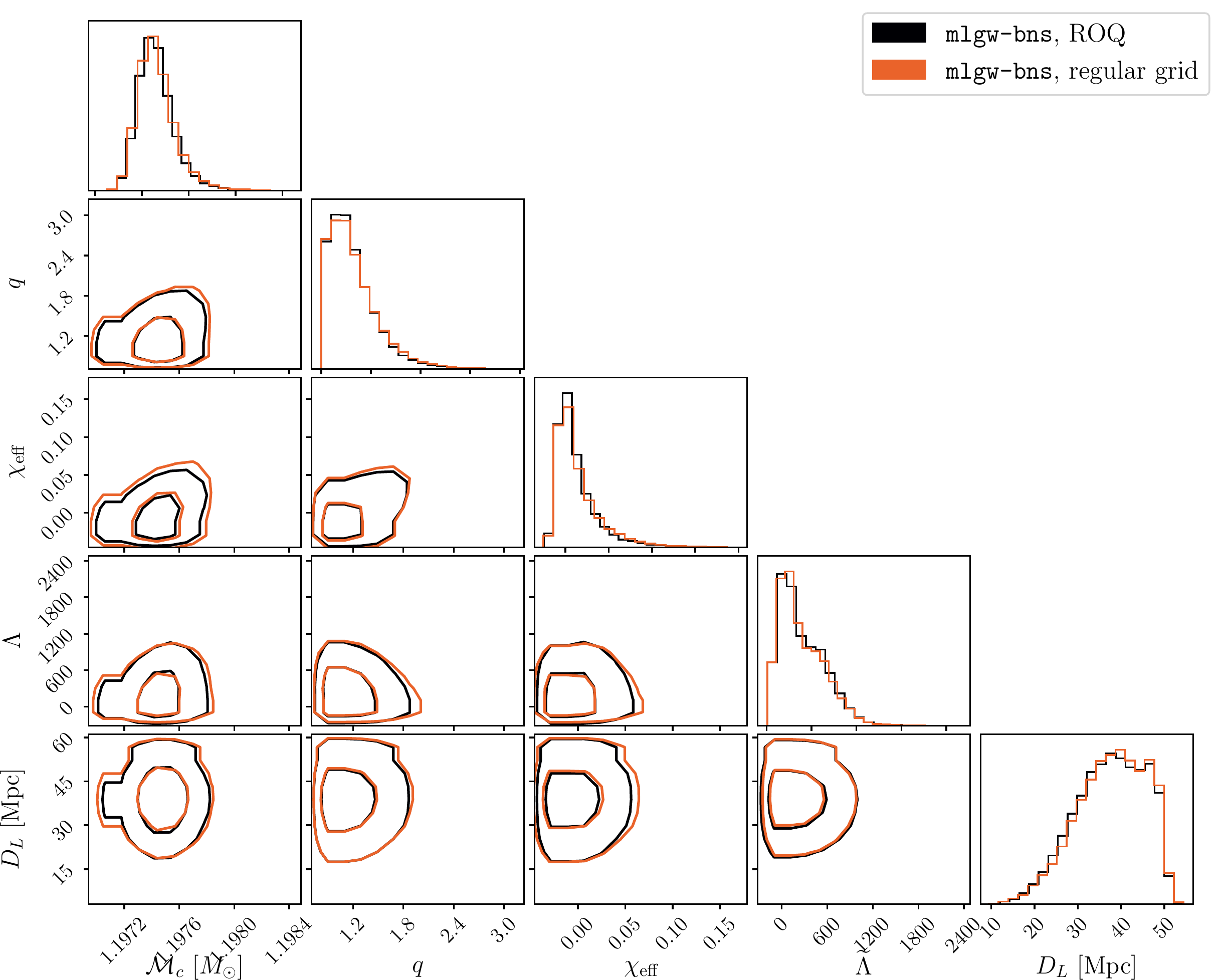}
\caption{Corner plot of the posterior distribution for selected parameters reconstructed for GW170817, in both cases with \mlgwbns, 
but when using an \ac{ROQ} technique or a full frequency grid evaluation.
The contours report the 50\% and the 90\% credibility regions.
}
\label{fig:corner_posterior_roq}
\end{figure*}

Using the settings discussed above, GW170817 is analysed with {\model} and {\teobspa} 
in order to compare performances and verify the consistency of the results.
The sampling employs 3000 live points, an evidence tolerance of 0.1, a maximum number of Markov-Chain Monte Carlo steps of 12000 and 5 auto-correlation times before accepting a point. We analytically marginalise over the coalescence time $t_c$ and phase $\phi_c$.

The two waveform approximants achieve compatible measurements, within the stochasticity of the sampler.
Figure \ref{fig:corner_posterior_mb_teob} shows the comparison between {\model} and 
{\teobspa} posterior distributions for selected parameters of interest.
We recover $\M={{1.1975}^{+0.0003}_{-0.0002}}~\Msun$,
the mass ratio is constrained to $q<2.07$ at the 90\% confidence level 
and the reduced tidal parameter corresponds to  $\tLam={{365}^{+522}_{-254}}$.
The recovered posteriors are consistent with previous similar 
studies~\cite{TheLIGOScientific:2017qsa,LIGOScientific:2018mvr,Abbott:2018wiz,Gamba:2020ljo,Breschi:2021wzr}.
Moreover, the two models recovered similar Bayes' factors ($\log\B \simeq 482$), 
and signal-to-noise ratios (${\rm SNR}={32}$),
validating the faithfulness of {\model} with respect to the training template in a realistic application.

We observe only a mild improvement in execution time for {\model} compared to {\teobspa}.
This is expected given the uniform frequency grid with $(f_{\text{max}} - f_{\text{min}}) \times T = (2000-23) \times 128 = 253056$ 
evaluation points. In fact, Fig.~\ref{fig:benchmarking-evaluation} shows that for this number of points 
the advantage in generating waveforms using {\model} is not enormous.
Significant speedups can instead be achieved by relying on grids smaller than $10^4$ points.
This naturally calls for the usage of compression techniques, capable of restricting 
the required number of frequency nodes used in computing the likelihood, the subject of the remainder of this section.

\subsection{Reduced order quadrature construction}\label{subsec:roq-con}

Reduced order modeling, which is referred to as \ac{ROQ}s in \ac{GW} astronomy when combined with discrete empirical interpolation techniques,
is a method of eliminating information redundancy present in sets of parametric functions (in our case, the gravitational waveforms as functions of the physical parameters of the binary system, such as masses and spins) when evaluated on a discrete set of points (in our case, the frequency grid). 
By selecting a small number of waveforms' ``basis elements'' and an equal number of discrete interpolation frequency points, \ac{ROQ}s are capable of dramatically speeding up both waveform evaluation and integrals involving them, such as the Wiener inner products (see Eq.~\eqref{eq:Wiener_ip}) entering the standard \ac{GW} likelihood.
This is achieved by sufficiently accurate -- and fast to evaluate -- \textit{interpolants}, built on a large training dataset.
In the context of \ac{GW} astronomy, early development and applications of \ac{ROQ}s to \ac{GW} searches were presented in Refs.~\cite{Field:2011mf, Caudill:2011kv}.
An extended mathematical analysis (notably, including convergence estimates) was presented in~\cite{Antil:2012wf}, while the construction of surrogate models using related techniques was pioneered in Ref.~\cite{Field:2013cfa}.
Applications to PE were introduced in~\cite{Canizares:2013ywa, Canizares:2014fya}, and the extension to precessing signals \ac{PE} was achieved in~\cite{Smith:2016qas}, also including many improvements such as mass-frequency partitions and an adaptive frequency sampling strategy. \ac{ROQ} acceleration of tests of \ac{GR} was considered in~\cite{Meidam:2017dgf}.
Most of the methods used in the aforementioned applications are implemented in the \texttt{GreedyCpp} code.\footnote{Available at: \href{https://bitbucket.org/sfield83/greedycpp}{bitbucket.org/sfield83/greedycpp}}
More recently, \ac{ROQ}s of precessing signals containing higher harmonics were presented in Ref.~\cite{Qi:2020lfr}, while Ref.~\cite{Smith:2021bqc} used \ac{ROQ} methods to demonstrate the feasibility of analysing \ac{BNS} merger signals detected by the next generation of ground-based detectors.
The interested reader may refer to~\cite{Antil:2012wf, Field:2013cfa, Smith:2016qas} for an introduction to the concepts used below.

Ref.~\cite{Qi:2020lfr} introduced a set of modifications in how the initial basis elements are constructed compared to previous literature, aiming at improving the efficiency of basis construction. 
The related algorithm was released in a public \texttt{python} package, labeled \texttt{PyROQ}.\footnote{Available at: \href{https://github.com/qihongcat/PyROQ}{github.com/qihongcat/PyROQ}}
We modified and generalised this algorithm, added numerical stability checks, restructured the software to make it more modular and easily usable with modern (typically \texttt{python}-based) waveform approximants.
Details of our algorithm, labeled \texttt{JenpyROQ},\footnote{Available at: \href{https://github.com/GCArullo/JenpyROQ}{github.com/GCArullo/JenpyROQ}} and GW170817 \ac{ROQ} interpolants construction are presented in Appendix A.
For the \ac{PE} analysis discussed above, we obtained a sufficiently accurate basis with 267 (10) linear (quadratic) basis elements, 
achieving a linear (quadratic) frequency axis reduction factor of 950 (25300).

\subsection{Parameter estimation with reduced order quadrature interpolation}\label{subsec:roq-pe}

\begin{table}[t]
\caption{Waveform generation ($\rm t_{\rm wf} $) and likelihood inner-products ($\rm t_{\rm ip} $) timings when using ROQs or a full frequency grid evaluation. 
We report results for both a single ($\rm N_{det} = 1$) and three detector network ($\rm N_{det} = 3$). 
The total likelihood evaluation time is simply $\rm t_{\rm tot} \simeq \rm t_{\rm wf} +\rm t_{\rm ip}$, since other likelihood operations costs are comparatively negligible.
The \ac{ROQ} approximation results in a PE speedup factor of 18 (12) in the one (three) detector case.}
\begin{tabular}{@{}lcccc}
\hline\hline
\multicolumn{5}{c}{Timings [ms]} \\
\hline
\hline
$(\rm N_{det}, \rm ROQ) $ & (1, no) & (1, yes) & (3, no) & (3, yes) \\
\hline
$\rm t_{\rm wf} $  & 69.8 & 2.2 & 69.8  & 2.2 \\
$\rm t_{\rm ip} $  & 15.3 & 2.5 & 45.9  & 7.5 \\
$\rm t_{\rm tot}$  & 85.2 & 4.7 & 115.7 & 9.7  \\
\hline\hline
\end{tabular}
\label{tab:ROQ_timings}
\end{table}

To predict the expected speedup on a \ac{PE} run using the \ac{ROQ} interpolants described above, it is sufficient to compute $\rm t_{\rm tot} = \rm t_{\rm wf} +\rm t_{\rm ip}$, where $\rm t_{\rm wf}$ indicates the waveform (Eq.~\ref{eq:emission-mode-decomposition}) generation time 
and $\rm t_{\rm ip}$ the evaluation time of the likelihood inner products (including interpolants evalutation): 
all other operations (e.g. detectors projection) are negligible compared to these two costs. 
Typical values for these times when using an \ac{ROQ} technique or a full frequency grid evaluation are reported in Table~\ref{tab:ROQ_timings}.
For a single detector, the predicted \ac{ROQ} speedup factor is $85.2~ \rm ms / 4.7~ \rm ms \sim {}18$.
For three detectors (the case of interest in our realistic application), the total speedup becomes: $115.7~ \rm ms / 10~ \rm ms \sim {}12$.
These numbers imply that when relying on {\mlgwbns} and an \ac{ROQ} scheme, the waveform evaluation cost is no longer the dominant one. 
For this reason, the expected \ac{PE} speedup (12) is a factor of three smaller than the waveform evaluation speedup (35) inferred from Fig.~\ref{fig:benchmarking-evaluation}.

We validate this by repeating the GW170817 analysis in the previous section, employing {\mlgwbns} both times but using either a \ac{GW} likelihood built with the \ac{ROQ} interpolants 
constructed above, or a standard likelihood computation.
We do not apply time-marginalisation in this case, since we have not interfaced the \ac{ROQ} formulation with the time-marginalised likelihood, hence we increase the values of sampler settings to avoid altogether any convergence issues.
We employ 5000 live points, an evidence tolerance of 0.1, a maximum number of Markov-Chain Monte Carlo steps of 12000 and 10 auto-correlation times before accepting a point.
We explore $t_c$ within the bounds [24.7, 25.0]~s, using a discretisation composed of 3000 points.
\ac{PE} results obtained with the \ac{ROQ} settings discussed above or with the standard likelihood are statistically indistinguishable, 
as shown in Fig.~\ref{fig:corner_posterior_roq}.
However, with 24 nodes comprising 2 Intel Xeon E5-2650v4 12x 2.20 GHz 12-Core CPU each, the sampling runtimes and relative speedup are: 
$\rm t^{ROQ=0}_{samp} / \rm t^{ROQ=1}_{samp} = 49\rm h 20\rm m / 4\rm h 14\rm m \sim 11.3$, in very good agreement with the predictions presented above.
Pre-sampling interpolant construction took 9~mins per detector with these settings.
Finally, we stress that the speedup resulting from the combination of {\model} and \ac{ROQ} will bear a more dramatic impact when applied to longer frequency axes.
For example, in the case of full inspiral-merger-postmerger \ac{BNS} signals analyses, with a lower frequency bound of ${\sim}5$ Hz and reaching up to ${\sim}8$ kHz,
applications of similar techniques will provide a speedup larger than three orders of magnitude compared to a uniform grid.

\section{Conclusions}
\label{sec:con}
In this work we have introduced \mlgwbns, a cutting-edge \ac{ML} surrogate waveform approximant in the frequency-domain for spin-aligned \ac{BNS} mergers, designed for applications to both current and future \ac{GW} detectors.
Our model is trained on highly accurate \teobspa~\ac{EOB} waveforms, faithfully represented with a fidelity larger than the accuracy of the baseline SPA model against the native time-domain \ac{EOB} model $(\mathcal{\bar{F}} \lesssim 10^{-5})$.
At the same time, thanks to several dimensional reductions steps, \mlgwbns {} delivers a remarkable increase in efficiency, achieving the fastest waveform generation among all the available approximants including comparable physical information.

By performing careful benchmark tests with varying frequency grids, we estimate a speed-up of $\sim 30$ with respect to \teobspa {}, when evaluated on frequency axes composed of less than $\sim 10^4$ points, which can reach up to $\sim 35$ for less than $\sim 10^2$ frequency points.
Combined with \ac{ROQ} techniques, an overall \ac{PE} acceleration of more than an order of magnitude is achieved for current \ac{BNS} analyses -- as we explicitly demonstrated re-analysing GW170817 using a reduced basis.
Thanks to the improved performance of our model, in our investigations the likelihood cost is no longer dominated by the waveform generation time, but by inner products' computations, making additional decreases in the evaluation time of our \ac{ML} model less relevant.
If the inner products' computation cost can be reduced in future \ac{PE} implementations, it will be important to explore further optimisations of the algorithm, such as tuning the number of \ac{PCA} components and the greedy downsampling reconstruction tolerance, or improving the hyperparameters selection procedure.

Analysis targeting next generation detectors' observations, such as \ac{PE} studies in the ET band, will benefit from even more dramatic improvements.
Since the number of empirical nodes will still be $O(10^2)$ even at high SNR~\cite{Smith:2021bqc}, well into the plateaux of Fig.~\ref{fig:benchmarking-evaluation}, \ac{ROQ} interpolants  will allow to keep the waveform generation cost almost identical to the one of current analyses. This in turn will lead to a waveform generation speed-up of \({\sim}50\), compared to a standard uniform grid when analysing a signal starting from $5 \rm Hz$.
Instead, given the extremely low overhead of our \ac{ML} model, the combined usage of \ac{ROQ} and \mlgwbns {} will provide a massive speed-up of more than \({\sim}10^3\) for the same configuration, without loss of accuracy.
Finally, thanks to the usage of PN-hybridisation techniques, our model can be evaluated from arbitrarily low frequencies, readily allowing analyses exploring the discovery potential of planned space-based deci-Hertz detectors. 

Other than exploiting fast \ac{PE} techniques, our \ac{ML} model can even enable them.
In fact, posterior sampling acceleration through \eg the application of Hamiltonian nested sampling \cite{Betancourt:2011rgh}, as well as forecasting with Fisher matrix studies, can be easily achieved thanks to the intrinsically differentiable architecture of \mlgwbns {}: a planned neural network upgrade is to yield not only the waveform polarizations \(h_{+, \times}\) but also their derivatives with respect to the parameters, \ie \(\partial h_{+, \times} / \partial \theta_i\).
The knowledge of gradients can be also exploited in template bank generation~\cite{Coogan:2022qxs}, allowing for a fast computation of a metric approximation for the match and for coverage of a large dimensional parameter space: our model will facilitate the generation of the first \ac{BNS} template bank including tidal effects.

In the future, the baseline model and the physics content of \mlgwbns {} will require improvements 
in order to meet the accuracy prerequisites of XG observatories.
While the simplicity in re-training \mlgwbns {} will allow it to remain up to date with future 
enhancements of tidal \ac{EOB} models (such as self-spin interactions, higher order tidal effects, dynamical tides), new challenges will be represented by the inclusion of: higher modes~\cite{Nagar:2020pcj}, precession~\cite{ Akcay:2020qrj,Gamba:2021ydi}, eccentricity~\cite{Chiaramello:2020ehz, Nagar:2021gss}\footnote{All these features are already implemented both in the native time-domain \teob {} model and in \teobspa {}, with the exclusion of eccentricity, only available in the time-domain waveform.} and a frequency-domain postmerger completion~\cite{Breschi:2022ens,Breschi:2022xnc}.
We leave such extensions of \mlgwbns {} to future work, but briefly discuss possible strategies to tackle them.
Higher order ($\ell>2$) modes break the simple dependence on the inclination angle \(\iota\) described by Eq.\eqref{eq:22-mode-polarizations}, requiring the modes to be reconstructed separately, with a corresponding slowdown in waveform evaluation. 
GPU acceleration~\cite{Thomas:2022rmc} could be employed to ameliorate this.
Precession effects could be immediately included relying on \ac{ML}-reconstruced higher modes, and subsequently applying a twisting~\cite{Schmidt:2010it, Schmidt:2012rh, Gamba:2021ydi} describing a generic spins dynamics.
Finally, eccentricity introduces modulations which make the time-to-frequency map non-monotonic: this prevents a straightforward application of \ac{SPA}, which we use to generate our training datasets. This problem could be cured by moving from \ac{SPA} to shifted uniform asymptotics \cite{Klein:2018ybm,Klein:2014gds}.

In summary, \mlgwbns{} enables an important leap towards feasible and accurate \ac{PE} with XG detectors, immediately providing a very efficient alternative to current \ac{EOB} \ac{BNS} models for present-day analyses.

\begin{acknowledgments}

  JT and SB thank Michela Mapelli for supporting this project and early discussions.
  GC thanks Hong Qi for discussions on \texttt{PyROQ} and Rory Smith, Carl-Johan Haster for useful insights on integrating detector calibration uncertainties with ROQ interpolants.
  MB and SB acknowledge support by the EU H2020 under ERC Starting Grant, no.~BinGraSp-714626.
  MB and RG acknowledge support from the Deutsche Forschungsgemeinschaft (DFG) under Grant No. 406116891 within the Research Training Group RTG 2522/1.
  GC acknowledges support by the Della Riccia Foundation under an Early Career Scientist Fellowship.
  GC acknowledges funding from the European Union’s Horizon 2020 research and innovation program under the Marie Sklodowska-Curie grant agreement No. 847523 ‘INTERACTIONS’, from the Villum Investigator program supported by VILLUM FONDEN (grant no. 37766) and the DNRF Chair, by the Danish Research Foundation.
  This research has made use of data, software and/or web tools obtained from the Gravitational Wave Open Science Center (https://www.gw-openscience.org), a service of LIGO Laboratory, the LIGO Scientific Collaboration and the Virgo Collaboration.
  LIGO is funded by the U.S. National Science Foundation. Virgo is funded by the French Centre National de Recherche Scientifique (CNRS), the Italian Istituto Nazionale della Fisica Nucleare (INFN) and the Dutch Nikhef, with contributions by Polish and Hungarian institutes.
  Computations were performed on {\scshape ARA}, a resource of Friedrich-Schiller-Universt\"at Jena supported in part by DFG grants INST 275/334-1 FUGG, INST 275/363-1 FUGG and EU H2020 BinGraSp-714626. Postprocessing was performed on the {\scshape Tullio} sever at INFN Turin.

  \noindent {\mlgwbns} is publicly available at:

  \href{https://github.com/jacopok/mlgw\_bns}{github.com/jacopok/mlgw\_bns}

  \noindent \texttt{JenpyROQ} is publicly available at:

  \href{https://github.com/GCArullo/JenpyROQ}{github.com/GCArullo/JenpyROQ}

  \noindent {\teobspa} is publicly available at:

  \href{https://bitbucket.org/eob\_ihes/teobresums/}{bitbucket.org/eob\_ihes/teobresums/}

  \noindent {\bajes} is publicly available at:

  \href{https://github.com/matteobreschi/bajes/tree/release/v0.3.0}{github.com/matteobreschi/bajes}.
  
  \noindent The Bayesian analyses presented in this work have been performed with {\bajes} version {\tt 0.3.0}, also available on \href{https://pypi.org/project/bajes/0.3.0/}{\scshape PyPI}.
\end{acknowledgments}

\newpage

\appendix

\section{Benchmarking procedure}
\label{sec:appendix_C}

In this section we detail the procedure used in order to compute the benchmarks shown 
in figure \ref{fig:benchmarking-evaluation}, and we draw a comparison to other existing
\ac{EOB} surrogates. 

The times we are measuring are quite short and susceptible to fluctuations, 
therefore we must average over several trials.
Further, the order in which tests are performed may affect the results, 
therefore we randomize the trials across several epochs.

We create a batch of test cases with all possible combinations of the following: 
\begin{enumerate}
  \item the \(N _{\text{app}}\) relevant approximants;
  \item the \(N _{\text{grid}}\) relevant frequency grid sizes \(n _{\text{freq}}\), chosen to be integer approximations of a logarithmically spaced grid;
  \item \(N _{\text{seed}}\) seeds for the parameters. 
\end{enumerate}

\begin{figure}[ht]
\centering
\includegraphics[width=.48\textwidth]{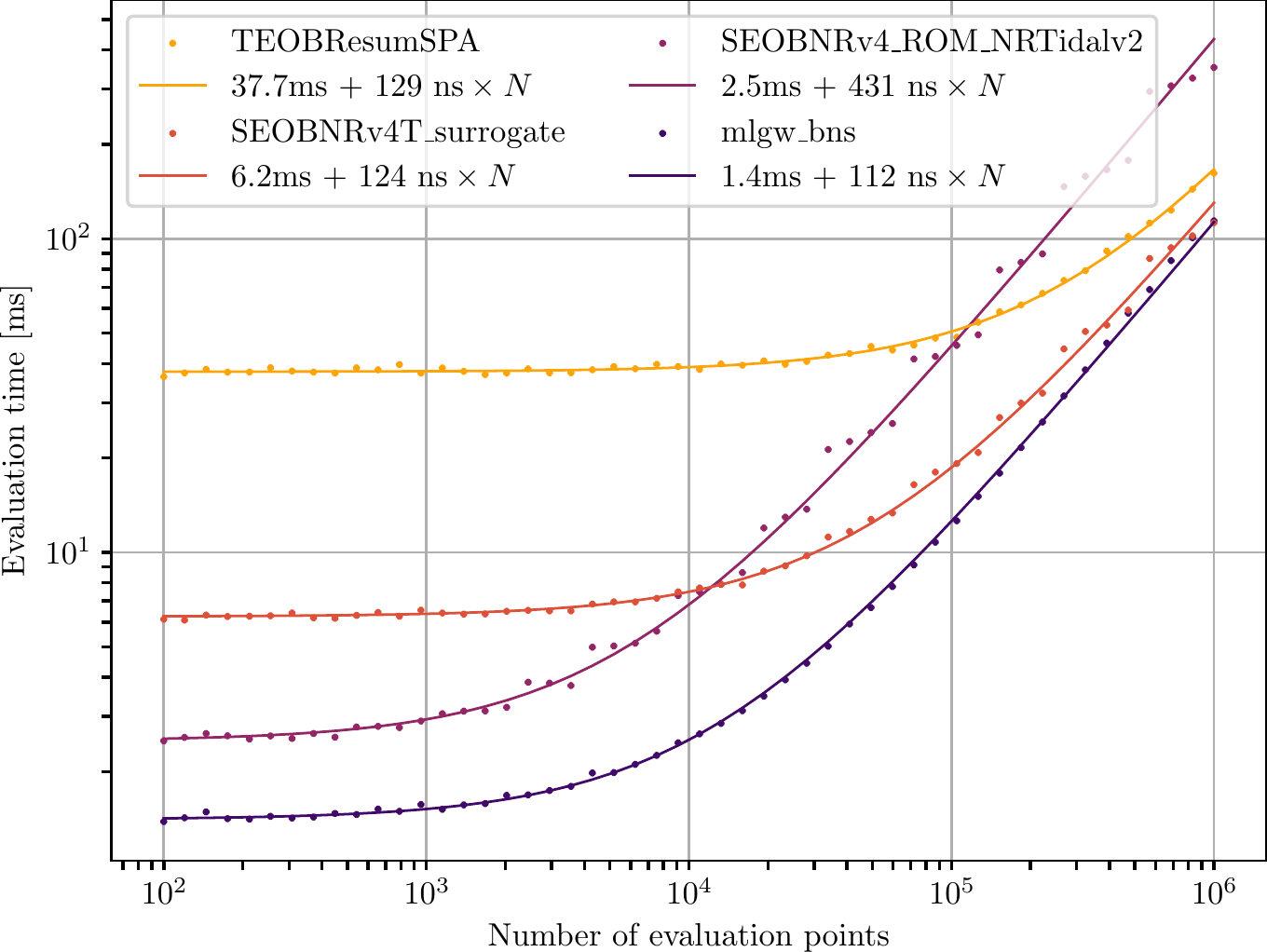}
\caption{Evaluation times for waveforms starting at 15Hz for several approximants. 
\texttt{SEOBNRv4\_ROM\_NRTidalv2} and \texttt{SEOBNRv4T\_surrogate} are called through their 
\texttt{LAL} python interfaces.}
\label{fig:benchmarking_evaluation_15hz}
\end{figure}

For each test case, we define a uniformly-spaced\footnote{As discussed in section \ref{sec:multibanding} this is not a good choice for \ac{PE}, but for the purposes of benchmarking the number of points in the frequency grid is the only relevant parameter.} frequency grid with \(n _{\text{freq}}\) points between \(f _{\text{min}}\) and 2048Hz.
In figure \ref{fig:benchmarking-evaluation} we chose \(f _{\text{min}} = 5\text{Hz}\), while in this section we choose \(f _{\text{min}} = 15 \text{Hz}\) in order to compare with other models, and the results can be seen in figure \ref{fig:benchmarking_evaluation_15hz}.
We then randomly generate a set of parameters, with the same procedure used for the generation of the model and with a different seed.
For the generation of figures \ref{fig:benchmarking-evaluation} and \ref{fig:benchmarking_evaluation_15hz} we use \(N _{\text{grid}} = 50\), \(N _{\text{seed}} = 20\), and run all tests for 10 epochs, shuffling them each time.

The \mlgwbns {} model used in figure \ref{fig:benchmarking_evaluation_15hz} is not the exact same one whose performance is discussed in the rest of the paper: the higher initial frequency allows us to use a smaller reduced frequency grid.
Besides this change, however, the training parameters are the same, and it was also verified to have mismatches \(\lesssim 10^{-5}\) with its training reference.

\section{Hyperparameters used}
\label{sec:appendix_B}

The hyperparameters used for the network whose performance is discussed in this work are as follows:
\begin{enumerate}
    \item the exponent \(\alpha\) for the principal components is set to \(0.37\);
    \item the network has two hidden layers with sizes 82 and 95 respectively;
    \item the activation function chosen is \(\tanh\);
    \item the L2 regularization parameter is set to \(10^{-4}\);
    \item the batch size for the Adam algorithm is set to 162;
    \item the initial learning rate is set to \(1.2 \times 10^{-3}\);
    \item the fraction of data kept for internal validation during the training is \(7.5\%\) (this refers to the internal validation step performed automatically by \texttt{scikit-learn});
    \item when the accuracy does not improve by more than \(2 \times 10^{-15}\) for 91 epochs the training stops. Note that this number refers to Euclidean distances across \ac{PCA}-reduced residuals, so it has no direct physical meaning.
\end{enumerate}

The downsampling indices are determined by training on a dataset of \(2^{10} = 1024\) waveforms, the \ac{PCA} is trained on \(2^{16} = 65536\) waveforms, the network is trained on \(2^{17} = 131072\) waveform residuals.
All these training datasets are independent, extracted from the same uniform distribution in the ranges of the parameters.

\section{ROQ construction}
\label{sec:appendix_A}

\subsection{PyROQ summary}

We first give a brief recap of the \texttt{PyROQ} algorithm presented in Ref.~\cite{Qi:2020lfr} (see their ``Algorithm 1''), to allow for an easier comparison with the applied software improvements and differences in our training strategy; we refer the reader there for a more detailed description.
In the first step of the \texttt{PyROQ} algorithm, a ``pre-selected'' dataset of waveform vectors (typically referred to as ``basis'') is constructed. 
This pre-selected basis is initially composed by waveform vectors evaluated at the corner of the parameter space.\footnote{Where the corners are not all the ones of the parameter space \(N\)-cube, but simply the two points consisting of the lower and upper bounds of each parameter. Also in our algorithm below, we chose to follow the same strategy, since we found that including all the \(N\)-dimensional cube edges of the physical parameter space leads to many repetitions which inefficiently increase the number of total basis elements.}
Corner elements are then augmented by randomly generating a waveform dataset of $10^5$ elements, and adding to the basis the element with the largest residuals after projection onto the basis. This is repeated until a user-specified tolerance is reached.
In a second step, the pre-selected basis is enriched by generating four increasingly larger datasets of $10^4,10^5,10^6, 5 \times 10^6$ elements, which might have different tolerance thresholds.
In each of these datasets, the element with the largest interpolation error is added onto the basis, iterating until all the elements of the dataset can be represented with a given accuracy.

\subsection{JenpyROQ summary} 

In this section we discuss our updated version of \texttt{PyROQ}, called \texttt{JenpyROQ}.
Compared to \texttt{PyROQ}, we allow for arbitrary enrichment cycles to be defined by the user, add a few strategies to avoid ill-conditioning of the interpolants construction (which is always monitored as a basic sanity check) and perform a restructuring of the code.
The latter consists of: imposing modularity at all stages; implementing MPI parallelisation, generalising the parameter handling (useful when dealing with flexible models with many more parameters than standard binaries, e.g. \ac{BNS} post-merger models~\cite{Breschi:2019srl, Breschi:2022xnc, Breschi:2022ens}); extending I/O management; adding a generic waveform class capable of interfacing with modern \texttt{python}-based models.

Regarding the algorithm itself (again, using as a reference template ``Algorithm 1'' of ~\cite{Qi:2020lfr}), in the pre-selection cycle, we complement the initial corner basis until the total basis size reaches either a given number of elements $N_{\rm pre}$ or a specified tolerance $\epsilon_{\rm pre}$ (which might be different from the final target tolerance $\epsilon_{\rm ROQ}$). 
At each pre-selection step $j$, with $j=1,...,N_{\rm pre}-2$, we generate $N_{\rm step}$ new random waveforms, among which we pick the element worst represented by projection on the current basis.
Later, in the enrichment steps, we adopt a more flexible scheme and perform an arbitrary number of enrichment cycles $N_{\rm cycles}$, each with arbitrary size $L^i$, number of outliers $N^i_{\rm out}$ and threshold $\epsilon^i$, with $i=1,..., N_{\rm cycles}$.
Finally, we test the constructed interpolants on $N_{\rm out}$ out-of-training datapoints.

By imposing a maximum $N_{\rm step}$ in the pre-selection phase (which can typically be guessed from rapid exploratory investigations), we keep the size of the pre-selection datasets ($N_{\rm pre} \times N_{\rm step}$) small, and so ensure that only a small number of evaluations are performed until the size of the basis is already large.\footnote{This does not apply to the cases where $\epsilon_{\rm pre}$ is reached, but in such a case only a small number of subsequent evaluations on larger datasets are expected if $\epsilon_{\rm pre} < \epsilon_{\rm ROQ}$, which we typically impose.}
Only after the pre-selected basis has a substantial number of elements, we compare against increasingly larger datasets in the enrichment steps. This way, most of the elements are already accurately represented, and no need of re-evaluating many comparisons arises, allowing us to keep a small memory footprint.
Furthermore, using a stricter relative tolerance ($\epsilon^i/\epsilon_{ROQ}$) on in the initial enrichment cycles ensures that an accurant interpolant is constructed with only a small number of computations.
Given the exponential convergence of the algorithm, such accuracy typically carries over subsequent much larger datasets at the true target tolerance.
The constructed interpolants are interfaced with the {\bajes} pipeline~\cite{Breschi:2021wzr}, following the likelihood formulation of~\cite{Smith:2016qas}.
Future planned developments include adaptive frequency sampling, on the lines of~\cite{Smith:2016qas} and of Sec.~\ref{sec:multibanding} of this work.

Finally, although the interpolants construction is always formally well-defined because the basis matrix in Eq.~16 of~\cite{Field:2013cfa} is always invertible,\footnote{Except for the trivial distance rescaling that may be factored out in \ac{ROQ} constructions, waveforms of binaries in \ac{GR} are always linearly independent. This discussion does not apply to models which include exact degeneracies among basis elements when varying the physical parameters.} in some cases the numerical inversion of the basis matrix can be ill-conditioned.
This happens when the algorithm finds the same empirical interpolant point more than once: they have to be unique to avoid double-counting of frequencies in the likelihood.
Ill-conditioning is more frequent for longer waveforms (corresponding to low-mass binaries), and it is caused by large disparities in the module of the basis matrix eigenvalues, implying a high conditioning number (when assuming a Euclidean norm), hence larger numerical instabilities.
Ill-conditioning appeared in early attempts to construct basis relevant to this work as soon as the segment length was increasing above $T = 128 \rm s$.
An easy solution we have found to this problem is simply to avoid applying a Gram-Schmidt procedure to the new basis elements (which are only strictly required to be independent, not orthonormal~\cite{Field:2013cfa}), using waveform vectors as basis elements as opposed to Gram-Schmidt residuals~\cite{Qi:2020lfr}. This avoids sparsity in the matrix construction, significantly decreasing the conditioning number.
For our current purposes, this simple fix was sufficient to ensure numerical stability. 
In case ill-conditioning is still present (a situation which we have not encountered yet in our explorations), currently the algorithm is forced to proceed, simply discarding the repeated interpolation point and switching to the interpolation point with the second worst residual.
We expect this latter procedure to slightly decrease the efficiency of the basis construction (i.e. to increase the number of elements required to reach the required precision), but given the exponential convergence of the algorithm, a small increase in the number of required elements does not appreciably affect \ac{PE} runtimes.
In the future, we plan to implement a more robust solution to the above issues by exploring modifications of our naive Gram-Schmidt algorithm, along the lines of what discussed in Appendix A of~\cite{Field:2013cfa} and their Refs.[66,67,68].
In any case, the matrix basis conditioning number is always monitored in our algorithm, and used to flag an ill-posed basis construction.

\subsection{Details of GW170817 interpolants construction}

We train the \ac{ROQ} basis on the full range of validity of {\model}, except for the chirp mass, which is only varied within the range $\mathcal{M}_c \in [1.1968,1.1988]$, wide enough to safely encompass the full posterior of GW170817.
This is not restrictive: the chirp mass can already be reliably estimated within an even tighter accuracy interval by low-latency \ac{GW} searches of \ac{BNS} signals~\cite{Biscoveanu:2019ugx}, hence a similar ``targeted'' basis could be constructed even for realistic, new data.
We set a tolerance threshold of $10^{-4} (10^{-6})$ for the linear (quadratic) basis, ensuring that the \ac{ROQ} interpolants are valid up to an SNR of $O(70-100)$, see e.g. Eq.~8 of Ref.~\cite{Purrer:2019jcp}.
We use a total of $4.1\times 10^5$ training datapoints, split between the pre-selection and the enrichment steps. 
A pre-selected basis is constructed using $N_{\rm pre} = 200 \, (10)$ elements for the linear (quadratic) case, and $N_{\rm step}=1000$ points at each step.
We set three enrichment cycles each composed of $[10^4, 10^5, 10^5]$ datapoints, $N^i_{\rm out}=0$ and a respective relative tolerance of [0.1,1.0,1.0].
The resulting bases are composed of 267 (10) linear (quadratic) elements, achieving a linear (quadratic) frequency axis reduction factor of 950 (25300).
We confirm the accuracy of the constructed interpolants by applying them to the reconstruction of $5 \times 10^5$ validation datapoints, finding less than $\rm 0.1\%$ outliers and none above the $10^{-3}$ threshold. 
Direct evaluation of $5 \times 10^5$ likelihood inner products confirmed the above tests, always showing errors smaller than the requested accuracy threshold.\\

The configuration file used to produce ROQ basis employed in this work is available at: \url{https://github.com/GCArullo/JenpyROQ/blob/main/config_files/config_MLGW-BNS_LVK_GW170817_release.ini}

\bibliographystyle{apsrev4-1}

\end{document}